# Convective and non-convective nature of Balearic meteotsunamis: a 50-year historical review

*A. Jansà[1], D. S. Carrió[1], C. Melo[2], R. Romero[1], B. Mourre[2,3] and V. Homar[1]*

[1]Meteorology Group, Department of Physics, University of the Balearic Islands, Palma, Spain.
[2]Balearic Islands Coastal Observing and Forecasting System, ICTS SOCIB, Palma, Spain
[3]Mediterranean Institute for Advanced Studies, CSIC-UIB, Esporles, Spain.

**Correspondence:** D. S. Carrió (diego.carrio@uib.es)

## ABSTRACT

Meteotsunamis, locally known as *rissaga* in the Balearic Islands, are significant sea-level oscillations induced by atmospheric disturbances, with amplitudes and frequencies comparable to seismic tsunamis. The port of Ciutadella (Menorca), is a recognized hotspot for meteotsunami occurrence. This study presents a 50-year analysis of 191 meteotsunami events recorded in Ciutadella between 1975 and 2025, focusing on the distinction between the convective and non-convective nature of the pressure disturbances that triggered the meteotsunami event and their implications for predictability and risk assessment. Events were classified using historical records, in-situ barographic and sea-level observations, satellite imagery, and lightning data. Statistical analyses explored relationships between atmospheric drivers, wave amplitude, amplification factors, and forecast performance. Results show that non-convective meteotsunamis are typically triggered by internal gravity waves, producing relatively regular and moderate pressure fluctuations. In contrast, convective events are associated with abrupt pressure jumps and tend to produce the most extreme sea-level oscillations—exceeding 3–4 meters in some cases, such as the catastrophic events of June 1984 and June 2006. Although non-convective events are more frequent, the most intense ones are predominantly convective. Nearly all meteotsunamis with amplitudes >220 cm are linked to convective disturbances. Forecasting skill varied by event type: A full-realistic high-resolution ocean-atmosphere modelling system (BRIFS) performed better for convective cases, whereas a targeted reduced-physics method (TRAM) was more accurate for non-convective scenarios. These results underscore the need to combine complementary modelling systems, as no single approach consistently performs well across all types of triggering mechanisms. This study provides the most extensive meteotsunami dataset for the Balearic Islands to date and offers a novel framework for understanding the contrasting dynamics and characteristics of both types of events. The results support the development of improved early-warning systems and more effective coastal risk mitigation strategies in vulnerable regions.

**Key words:** meteotsunami, rissaga, Ciutadella, pressure disturbances, convection, convective systems

## 1. Introduction

Meteotsunamis are large sea-level oscillations that occur within the same frequency band as seismic tsunamis, but with a meteorological origin (Monserrat et al., 2006). A range of meteorological phenomena can trigger meteotsunamis, including atmospheric gravity waves, rapid pressure jumps or surges, frontal passages, squalls and hurricanes (Vilibić et al., 2021). Meteotsunami formation generally involves three key stages: (1) the occurrence of a primary atmospheric disturbance, (2) a direct marine response (e.g., large marine waves), and (3) amplification of this response through specific mechanisms. Amplification can result from processes such as *Proudman* (or Greenspan) resonance, shoaling, and resonance within harbors, bays or inlets (Monserrat et al., 2006).

Several coastal regions worldwide are particularly prone to meteotsunamis, due to the frequent occurrence of suitable atmospheric disturbances and the high efficiency of local amplification mechanisms. Among these regions, the Balearic Islands in the Western Mediterranean, and, particularly, the port of Ciutadella, in Menorca, stand out as one of the most active hotspots for meteotsunami occurrence (**Fig. 1**).

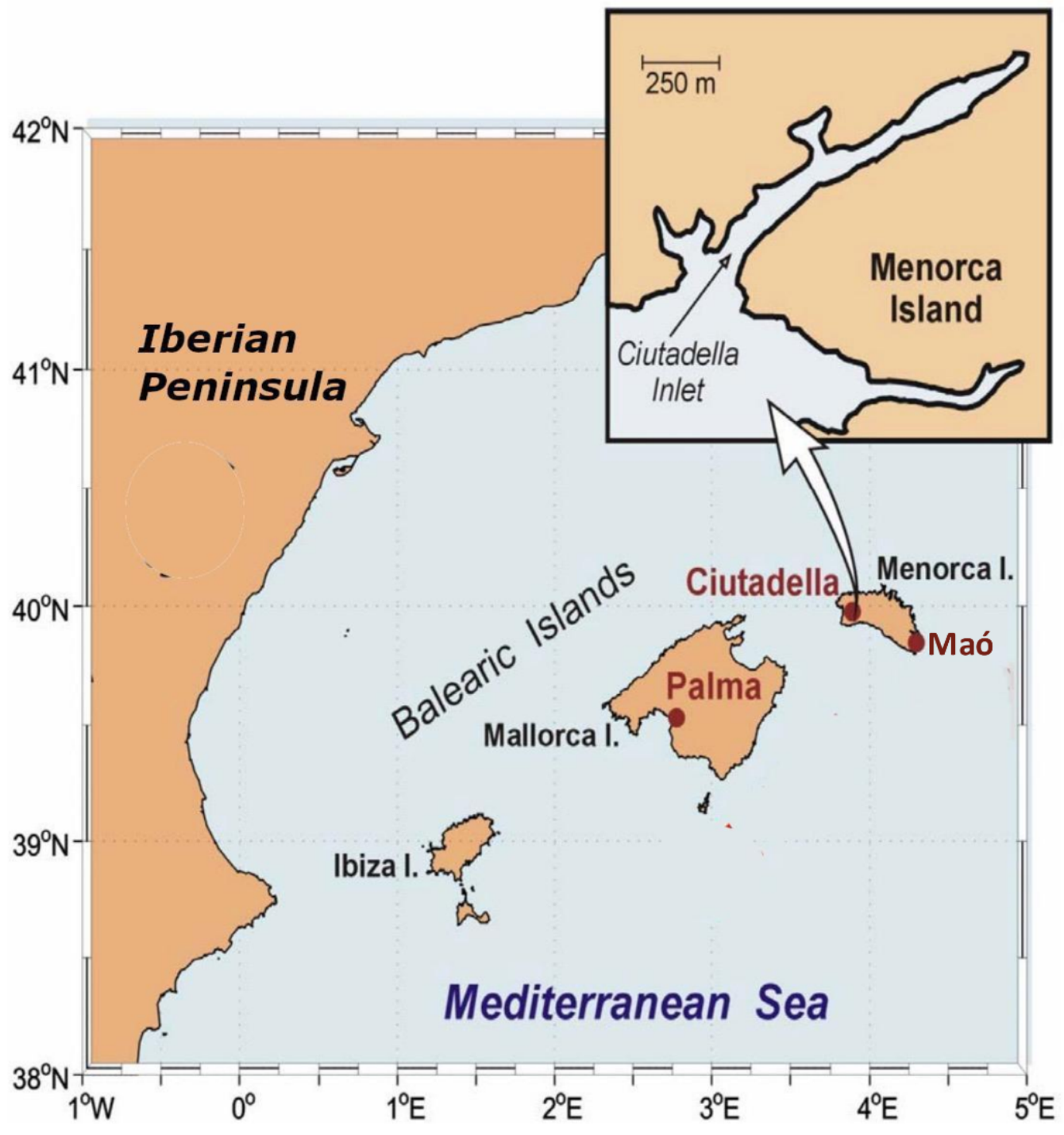

**Fig. 1**. *Location of the Balearic Islands in the Western Mediterranean. The map highlights Menorca Island and the port of Ciutadella, as well as the main locations mentioned in the text.*

Meteotsunamis in the Balearic Islands are locally known as *rissaga*. Despite the high local relevance of these events, a comprehensive climatology on *rissaga* events, particularly in Ciutadella, remains lacking, due to the relatively late introduction of instrumental monitoring, frequent gaps in observational continuity, and occasional data losses. As a qualitative climatological estimation, *rissaga* events with an amplitude (defined as the difference between the min and max of the wave height) of about 1 m occur in Ciutadella at least once per year. Indeed, more than 100 such events have been documented in less than 50 years. In contrast, *rissaga* events in other Balearic ports and inlets are significantly less frequent.

Several recent studies have identified specific types of atmospheric perturbations as the primary triggers of meteotsunamis. For example, Sibley et al., (2021) reported a meteotsunami linked to a convective rear-flank downdraft, while Pellika et al., (2020) associated Baltic Sea meteotsunamis with convective systems and lightning activity. Heidarzadeh and Rabinovich (2021) noted a partial contribution of typhoons to meteotsunami generation. In some regions, such as the Great Lakes, most documented meteotsunamis have been associated with convective storms (Bechle et al., 2016). In contrast, meteotsunamis in the Mediterranean are most frequently attributed to internal gravity waves that generate rapid atmospheric pressure oscillations (Ramis and Jansà, 1983; Vilibić et al., 2020).

In the Balearic Islands, and especially in Ciutadella, most meteotsunamis are triggered by relatively regular and moderate surface atmospheric pressure oscillations, typically identified as signatures of internal gravity waves. Spectral analyses, even from early analogical records, provided clear evidence of this mechanism (Ramis and Jansà, 1983). The later deployment of digital micro-barographs at several sites provided more robust evidence, even allowing for detailed characterization of these waves (Ramis and Monserrat, 1991; Monserrat et al., 1991; Monserrat and Thorpe, 1992; Villalonga et al., 2024).

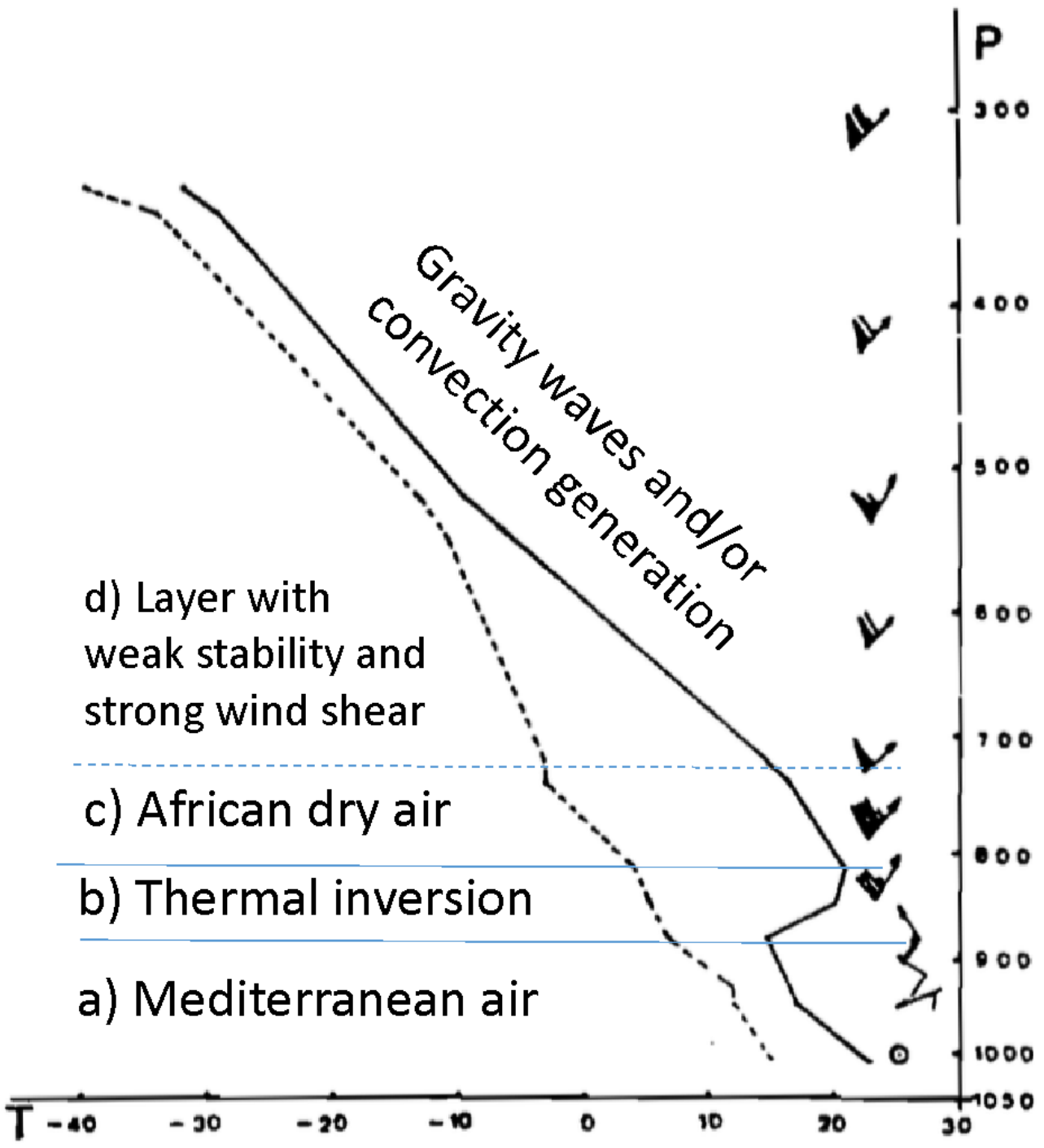

**Fig. 2**. *Vertical atmospheric profile (Stüve diagram) on 2-July-1981 (Adapted from Ramis & Jansà, 1983)*

Ramis and Monserrat (1991), Monserrat et al., (1991) and Monserrat and Thorpe (1992) used a theoretical three-layer model to describe the vertical atmospheric conditions appropriate to develop and maintain large-amplitude gravity waves. This framework proved to be compatible with the description of the synoptic meteorological conditions found to be empirically favorable to rissaga formation in the Balearic Islands (Ramis and Jansà, 1983; Jansà et al., 2007, Šepić et al., 2016; Jansà and Ramis, 2021). The favourable synoptic meteorological conditions typically include an upper-level cut-off low or trough centered to the southwest of the Iberian Peninsula, or over it, producing strong south-westerly winds over the Balearic Islands; a very warm and dry African air mass flowing, at low-medium levels, into the Western Mediterranean; and a weak sea-level low-pressure center in the Mediterranean region, accompanied by easterly surface winds.

These synoptic conditions comprise, from surface to mid-upper tropospheric levels, the following elements in the vertical temperature, humidity and wind profiles (**Fig. 2**):

a) A low-level layer of stable, warm and humid Mediterranean air.

b) A thermal inversion or a shallow, strongly stable layer around 800–900 hPa.

c) A layer of very warm, dry African air, just above the inversion/stable layer.

d) A temperature profile that decreases sharply above the previous layer, accompanied by strengthening southwesterly winds.

The resulting vertical temperature gradient and wind shear yields low Richardson numbers in the medium-upper layer, occasionally approaching conditional instability and favouring the gravity waves generation. When the vertical profile is favorable to *rissaga* events (**Fig. 2**), gravity waves would mainly form at mid-to-upper levels, around 500-700 hPa. These *internal* gravity waves can propagate both upwards and downwards in the atmosphere, becoming ducted waves, between the tropopause and the inversion top layer, travelling relatively long distances. Downward-propagating gravity waves induce oscillations in the top of the inversion layer, which is crucial because the strong thermal contrast between the warm air above and cooler air below amplifies the associated surface pressure oscillations through hydrostatic effects (**Fig. 3**).

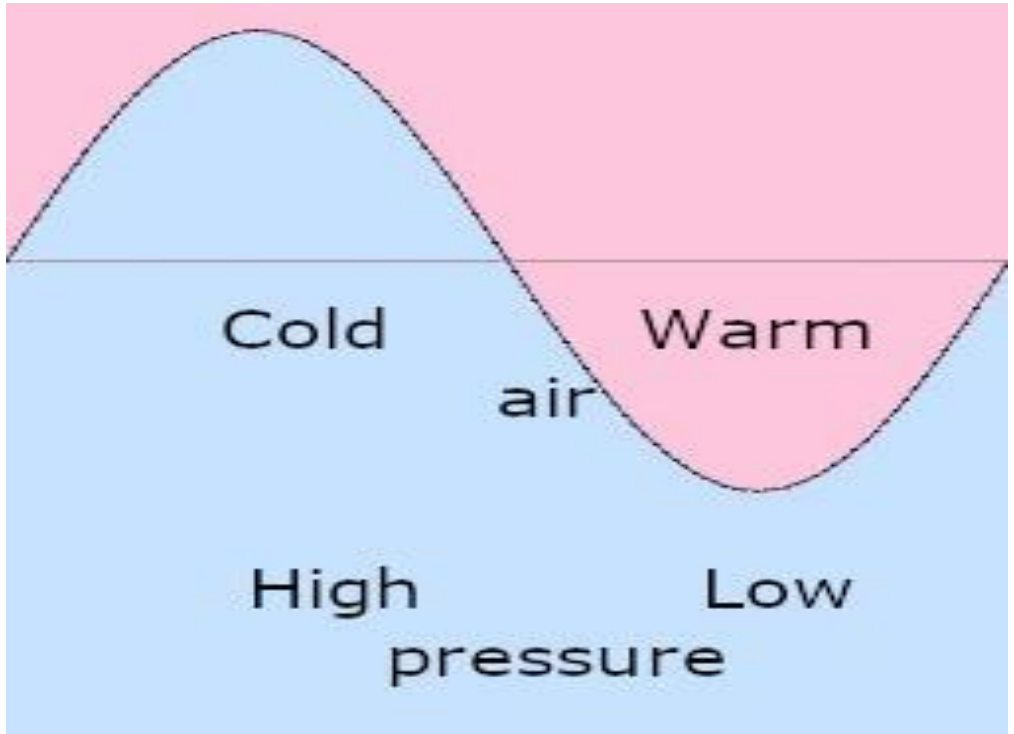

**Fig. 3**. *Schematic representation of a gravity-wave affecting the thermal inversion, with indication of warm/cold anomalies and the resulting pressure perturbations via the hydrostatic effect.*

However, gravity waves alone cannot account for all Mediterranean meteotsunamis. Pressure jumps or rapid pressure oscillations of convective origin have also been identified as responsible for some Mediterranean meteotsunamis. In particular, the catastrophic Ciutadella events of June 1984 and June 2006 (Jansà, 1986; Jansà et al., 2007) were associated with abrupt pressure changes embedded within well-developed convective systems.

The catastrophic *rissaga* of 21 June 1984, with an amplitude of near 4 meters, provided new insights about the nature of atmospheric disturbances responsible for large amplitude Balearic *rissaga* events. During this episode, an exceptionally strong pressure jump, exceeding typical variations associated with gravity waves, was observed. This signal coincided with the arrival of a squall line, identified as the gust front of a convective system clearly depicted through satellite imagery (Jansà, 1986). The passage of the squall line over Ciutadella coincided precisely with an extreme sea-level oscillation, whose amplitude was estimated (not measured) at 3-4 meters.

Convection is linked to conditional instability, while gravity waves require stable conditions, even in saturated air. Despite this contrast, both phenomena can coexist under marginal stability combined with strong vertical wind shear. In such cases, large gravity waves may locally lift air to the level of free convection following saturation, allowing elevated convection to develop (Markowsky and Richardson, 2010) despite the presence of significant Convective Inhibition (CIN) associated with the thermal inversion (**Fig. 2**). Conversely, vigorous convective updrafts are capable of generating gravity waves themselves. As a result, gravity waves and convection may appear simultaneously or in sequence, a dynamic interaction noted in several studies about gravity waves and convection (Uccellini, 1975; Bosart and Sanders, 1986).

Both subjective or instrumental observations, supported by satellite imagery, have shown that the arrival of singular pressure disturbances and/or particular and dense cloud structures over Ciutadella often coincides with large and singular sea-level oscillations (Jansà and Ramis, 1989, Jansà and Ramis, 1990, Jansà et al., 2001). During the catastrophic *rissaga* of 15 June 2006, again with an amplitude of near 4 meters, a singular pressure perturbation of convective origin was detected. The arrival to Ciutadella of the pressure jump and the extraordinary sea-level oscillation occurred simultaneously (Jansà et al., 2007).

Interestingly, the synoptic meteorological patterns observed in classic gravity-wave cases—such as 2 July 1981—are often similar to those seen in convective cases involving singular pressure jumps, such as the destructive events of 21 June 1984 and 15 June 2006. In this context, the present study addresses an important question concerning the two classes: Are *rissaga* events involving convection inherently more severe than those triggered solely by gravity waves?

The main objective of this study is to analyse the specific characteristics and impacts of convective versus non-convective *rissaga* events in Ciutadella. Following the terminology proposed by Rabinovich (2020), one could refer to these two types as "bad weather" and "good weather" *rissagas*, respectively. We examine their relative frequency and key phenomenological differences, including wave amplitude, magnitude of rapid pressure changes, and amplification factor linking offshore disturbances to local sea-level oscillations. In addition, the predictability of convective and non-convective *rissaga* events is also assessed.

To achieve these goals, we compiled a comprehensive dataset of *rissaga* events in the Ciutadella harbor, spanning 50 years from 1975 to 2025. This dataset, provided in Appendix 1, represents the most extensive archive of its kind for the region and includes detailed descriptors for each documented event.

This paper is structured as follows: Section 2 describes the data sources and the methodology used to build the historical dataset of *rissaga* events and classify them according to their convective and non-convective nature. Section 3 provides illustrative examples of different types of *rissaga.* Section 4 offers the main results regarding *rissaga* characteristics and predictability challenges. Finally, Section 5 presents the conclusions of this study.

## 2. Observational data and methodological approach

Beyond a few earlier references, systematic documentation of *rissaga* events began in 1975, coinciding with the start of dedicated research on the phenomenon in the Balearic Islands and the identification of rapid surface pressure oscillations as the key trigger (Ramis and Jansà, 1983; Jansà and Ramis, 2021). Earlier studies had already suggested a similar link. Fontserè (1934) reported rapid changes of atmospheric pressure with large sea-level oscillations in the port of Barcelona, and noted that the largest pressure changes often coincided with convective (thundery) conditions.

### 2.1. *Rissaga* events catalogue

We compiled 50 years (1975-2025) of records of *rissaga* events from multiple sources. The core dataset was built from long-term research efforts on *rissaga* cases (Jansà and Ramis, 2021) and consists of an extended list of events, each with succinct descriptive details. This basic list was complemented with information from the literature, particularly studies based on dedicated observational campaigns. Several instrumental deployments, including in-situ digital measurements, some of them taken within the port of Ciutadella, were carried out under the direction and partial funding of the *University of the Balearic Islands* (UIB), with international collaboration and support from the *Spanish Institute of Oceanography* (IEO). These campaigns took place during several spring-summer seasons, between 1988 and 1992. Papers reporting on these campaigns have allowed us to add events and details to our list (Monserrat et al., 1991; Monserrat and Thorpe, 1992, Gomis et al., 1993; Garcies et al., 1996). A later campaign carried out during 2007-2009, financed by *Ports of the Balearic Islands' Agency* (hereafter, *Ports IB*), also contributed valuable data. In order to complete our list of *rissaga* cases, unpublished case lists from official organizations (AEMET[1], UIB or SOCIB[2]) were consulted and used in this study. Nonetheless, a significant information gap remains between 2000 and 2003.

The available information is highly heterogeneous among cases. For some events, only a brief note and an approximate estimate of the wave amplitude were available; others included more detailed observations, such as sea-level and pressure data, satellite imagery, or lightning detections. The dataset is neither exhaustive nor homogeneous, and some events may have been missed. In many cases, there was insufficient information to determine whether the triggering disturbance was convective or non-convective.

The final dataset includes 191 events in total. Only events with amplitudes exceeding 60 cm were retained. Note that this threshold is close to the threshold considered by AEMET in their *rissaga* warnings, which is 70 cm. When sea-level oscillations above the threshold amplitude spanned multiple consecutive days, they were grouped as a single event.

For each case, we recorded the following parameters, whenever available:

1) Maximum sea-level oscillation amplitude observed in the port of Ciutadella.
2) Maximum pressure variation during the event.
3) Amplification factor, obtained by dividing 1) by 2).
4) Amplitude of the maximum sea-level oscillation forecast by the TRAM system, developed by UIB.
5) Amplitude of the maximum sea-level oscillation forecasted by the BRIFS system, operated by SOCIB.

Appendix 1 presents the complete event list, including date (year, month, day(s)) and the five key parameters described above. Missing values are denoted as "NA" (Not Available).

**2.2 Amplitude estimation and records**

When an inlet oscillates in its fundamental mode, the maximum wave amplitude occurs at the end of the inlet, while the minimum is found at the mouth of the inlet (Rabinovich, 2009). Therefore, the reported amplitude of a *rissaga* event depends on the location of the measurement or estimate. In fact, in Ciutadella most instruments were historically installed in an intermediate point of the inlet rather than at the end, implying that instrumental records may underestimate the true maximum amplitude. Visual estimates, on the other hand, often produced significantly higher amplitude values, likely due to observation at the resonance-prone end of the inlet.

Most of the *rissaga* events, particularly the old cases, in the absence of in situ observational instruments, relied mainly on witness information. A particularly reliable source was Joan Seguí, who worked for Ports IB and oversaw operations at the port of Ciutadella for many years. Complementary atmospheric pressure data for this period were obtained from instruments located outside the port of Ciutadella, such as those in Palma and Maó (**Fig. 2**).

Several dedicated field campaigns, described previously, provided discontinuous instrumental data, including amplitude. Continuous monitoring began in 2007, when *Ports IB* installed a permanent meteorological and oceanographic station inside the port of Ciutadella. The station, designed jointly with AEMET, records sea-level and atmospheric pressure every 30 seconds, a temporal resolution sufficient to capture peak sea-level oscillations and associated pressure changes during *rissaga* events.

---

[1] *State Meteorological Agency of Spain*
[2] *Balearic Islands Coastal Observing and Forecasting System*

Unfortunately, a substantial portion of the dataset from this station was lost, due to technical issues. Nevertheless, graphical outputs of several *rissaga* events had been stored prior to data loss and were included in the present analysis.

In September 2014, the SOCIB deployed a water bottom pressure recorder close to the *Ports IB* station. This instrument allows the inference of high-frequency sea-level variability, and therefore of meteotsunami amplitudes. Except for two data gaps (October 2018-July 2019 and July-October 2022), this series provides a valuable long-term record of *rissaga* amplitudes in Ciutadella. At present, dense observational networks of pressure and sea level across the Balearic Islands support the identification of new events and enhance the understanding of rissaga mechanisms, not only in Ciutadella but also in other locations throughout the region (Villalonga et al., 2024; Ramos-Alcántara et al., 2025).

### 2.3 Atmospheric pressure variations

The second key variable of interest is the magnitude of the rapid pressure oscillations and/or singular pressure jumps, when present. For early events, when no barometers were installed in Ciutadella, we relied on analogical records from nearby stations, such as the AEMET offices in Palma and Maó. These records were used as proxies to estimate the shape and magnitude of pressure variations in Ciutadella, under the assumption that pressure perturbations propagate without a fundamental change (Villalonga et al., 2024). In addition, high-quality in situ data were collected during the field campaigns of 1988-1992.

Since 2007, detailed and nearly continuous pressure variation information has been available in Ciutadella. Initially, part of this information was only accessible graphically and values were extracted directly from plots. From 2014 onward, however, digital pressure records from the SOCIB barometer enabled objective quantification of pressure variations. Accordingly, the pressure data presented in Appendix 1 are based on continuous digital records only after 2014.

### 2.4 Classification of *rissaga* events: convective vs non-convective

To analyze differences in the behavior of *rissaga* events, it is first necessary to discriminate between convective and non-convective cases. Several sources of information are useful for this purpose, including pressure records, satellite imagery, and lightning data.

#### 2.4.1 Diagnostic features in pressure records

There are several ways to distinguish *rissaga* events with convective and non-convective meteorological origin. Pressure evolution provides a first indication of the event type. In the absence of convection, trains of gravity waves typically generate relatively regular and moderate pressure oscillations. This was the case during the major *rissaga* of 2 July 1981 (Ramis and Jansà, 1983), when analogical barograms from Palma showed no abrupt pressure jumps. The maximum sea-level oscillation in Ciutadella reached 2 meters, approximately (**Fig. 4**).

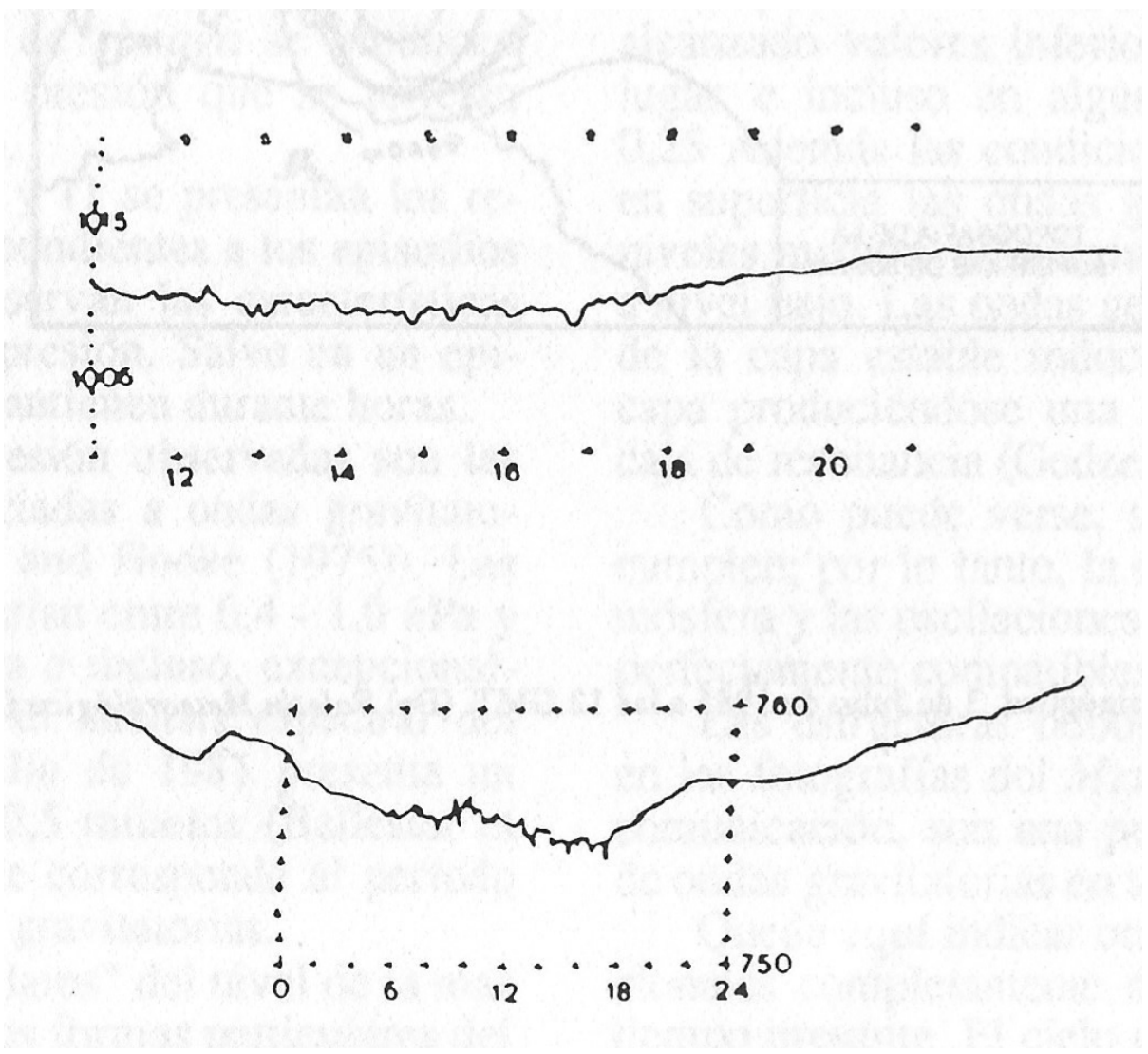

**Fig. 4**. *Two analogical barograms from AEMET in Palma, during 2 July 1981. The upper record captures pressure variations with higher temporal resolution than the lower one. Time (hours) is shown on the x-axis, pressure on the y-axis (hPa in the upper barogram; mmHg in the lower panel). Adapted from Ramis and Jansà, 1983.*

By contrast, convective systems often produce abrupt and singular pressure changes. **Fig. 5** shows the pressure evolution in Palma from 19 to 21 June 1984. Early on the 21$^{st}$, as a convective system passed over Mallorca and Menorca, the barogram recorded an abrupt and singular pressure change, characterized by a sharp rise followed minutes later by a rapid fall. Just when the main pressure perturbation arrived at Ciutadella, an extraordinary and catastrophic sea-level oscillation occurred (Jansà, 1986). It should be noted that the largest pressure variation at 04 UTC was accompanied by other high-frequency pressure changes during the three-day period. Although an abrupt and pronounced pressure jump often suggests convective activity, it does not guarantee it, as a particularly strong solitary gravity wave can produce a similar signal. Conversely, even in an environment able to become convective, moderate and regular pressure oscillations may still indicate the presence of gravity waves. As noted earlier, convective activity and the presence of gravity waves can coexist within the same day or event (see **Fig. 5** and **Fig. 8**).

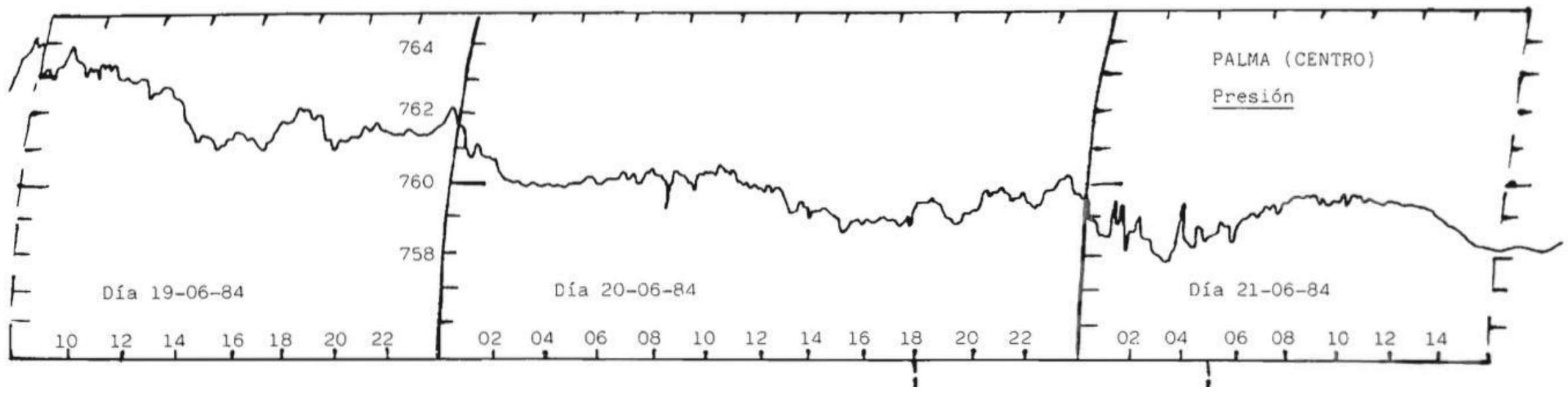

**Fig. 5**. *Three-day barogram recorded at AEMET in Palma, from 19 to 21 June 1984. In the early hours of the 21st, a convective pressure jump coincided with the onset of an extraordinary and catastrophic rissaga event in Ciutadella harbour (Jansà, 1986).*

As noted earlier, an important and singular pressure jump or pressure oscillation may indicate a convective situation, but this is not necessarily the case: such a signal can also result from the passage of a large, singular, solitary gravity wave. Distinguishing between the two is often challenging, and pressure variations alone are neither the only nor the most reliable criterion for classifying a *rissaga* event as convective or non-convective.

### 2.4.2 Remote-sensing indicators of convection

Beyond pressure evolution, satellite imagery provides an additional tool to discriminate between convective and non-convective pressure disturbances. Notably, *rissaga* occurrence requires at least partly cloudy skies (Jansà and Ramis, 1989, 1990), since saturated air is inherently less stable than dry air.

On the synoptic scale, *rissaga* situations are often associated with the formation of large "comma-shaped" cloud systems, which can be interpreted as *Warm Conveyor Belts*, according to Browning (1997). These structures promote the ascent of very warm African air, leading to saturation of the air mass, while strong south-westerly upper-level winds develop over the Mediterranean (**Fig. 6**).

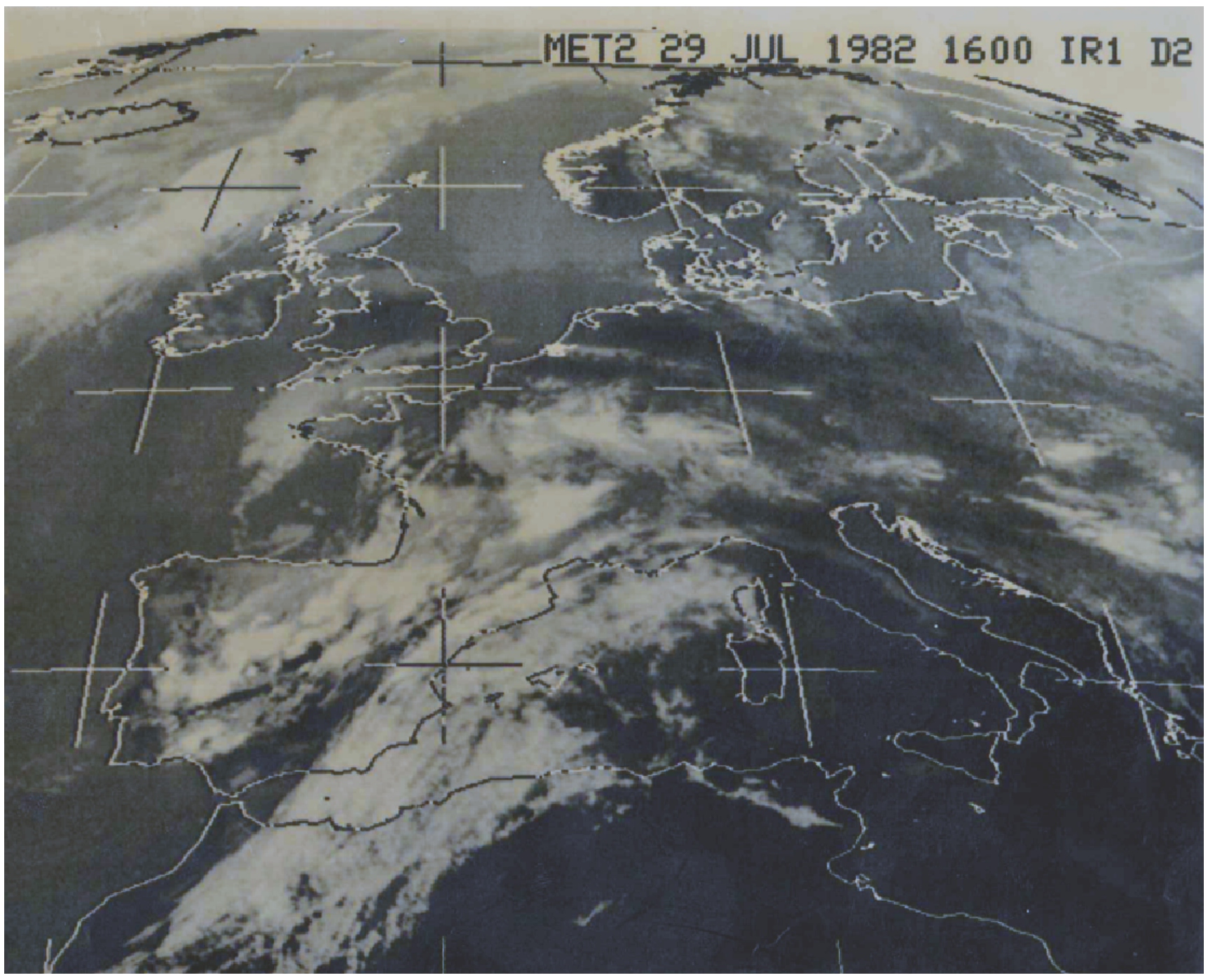

**Fig. 6**.- *Comma-shaped cloud organization, identified as a Warm Conveyor Belt. No convection is visible at the moment of the picture, suggesting this case is a candidate to be classified as a non-convective rissaga event (Meteosat image, 29 July 1982, at 0700 GMT; IR channel)*

An almost continuous cloud pattern (**Fig. 6**) or a reticular cloud cover (**Fig. 7**) is indicative of a non-convective situation. By contrast, when convection is present, cloud systems appear as bright cloud cores, with well-defined boundaries.

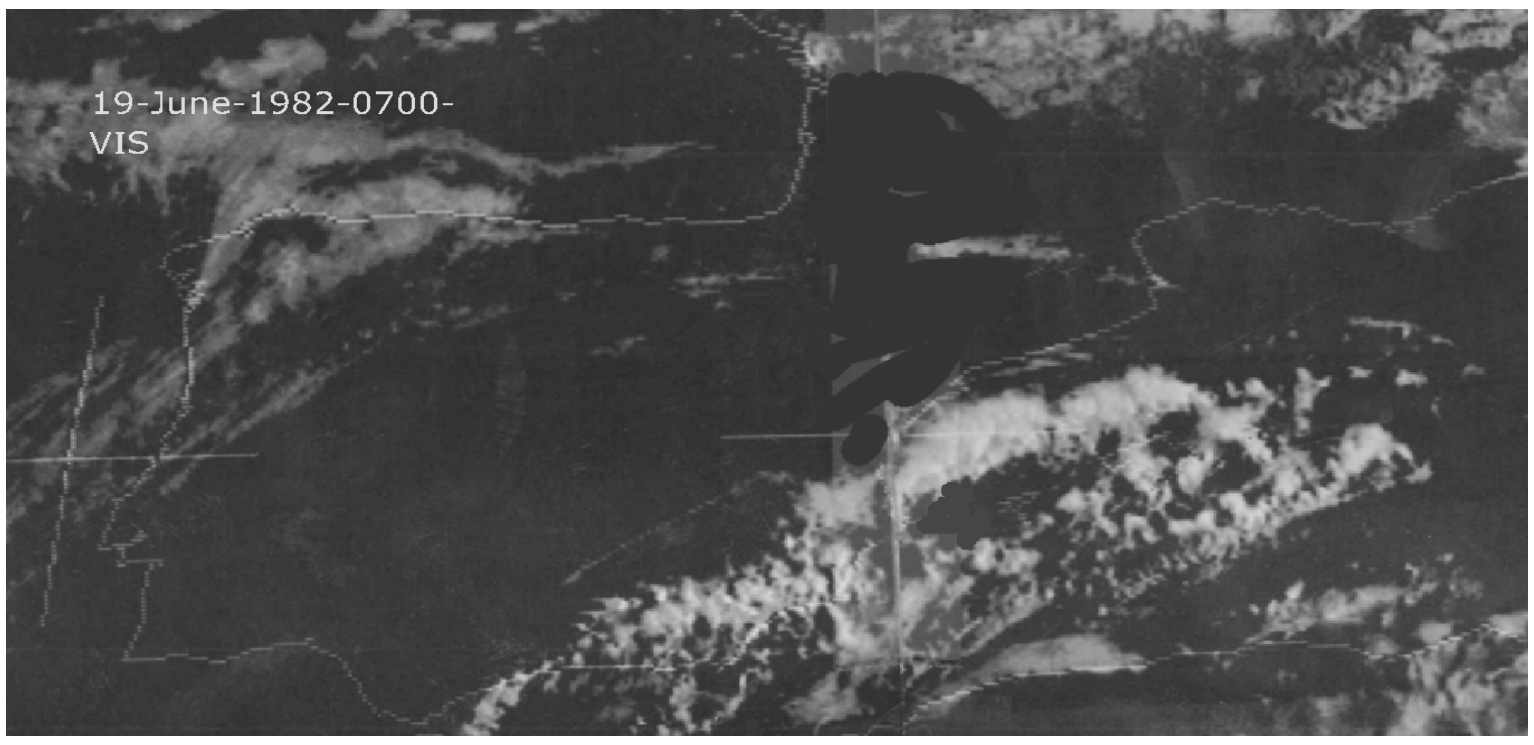

**Fig. 7**. *Meteosat visible image (19 June 1982, at 0700 GMT) showing gravity-wave crests and troughs as bright-dark bands, transverse to the main upper flow.*

The comma cloud can be more or less compact and continuous (**Fig. 6** and **Fig. 7**). When it exhibits a banded texture, that is, bright/dark lines oriented transverse to the upper-level main air-flow, without embedded convective systems, then we can define the perturbation as a non-convective.

On the contrary, the presence of bright, compact, rounded and well-defined cloud structures in the satellite imagery is indicative of convective systems, and the associated event can be classified as convective. Note that a situation may evolve from non-convective, with or without gravity waves, to convective.

The catastrophic *rissaga* of 15 June 2006 (Jansà et al., 2007) provides a clear example of a transition from non-convective to convective situation in the Balearics. Early in the day, stable vertical temperature gradients favoured the development of gravity waves, clearly visible south of the islands (**Fig. 8a**). By contrast, later in the day, the environment evolved into a convective scenario dominated by organized convective systems (**Fig. 8b**). High-resolution atmospheric pressure records from Palma and Maó (AEMET) illustrates the same idea: small rapid pressure oscillations (0,5-1,5 hPa) are visible most of the time, during the long-lasting "gravity waves phase", **Fig. 8a** and **Fig. 8c**, while, at the end of the day, powerful convective systems have appeared in the zone. In the Balearics, the arrival of a large convective system is simultaneous with a large pressure jump of up to 7 hPa, which is recorded in Mallorca and Menorca. Likewise, when this pressure jump was just arriving at Ciutadella, an extraordinary and damaging sea-level oscillation impacted the port. This event is classified as convective because the convective phase, although shorter than the preceding gravity wave phase, proved far more significant, with a big impact. In general, we agreed to define an event as convective if a "convective phase" is observed during an event, regardless of its relative duration.

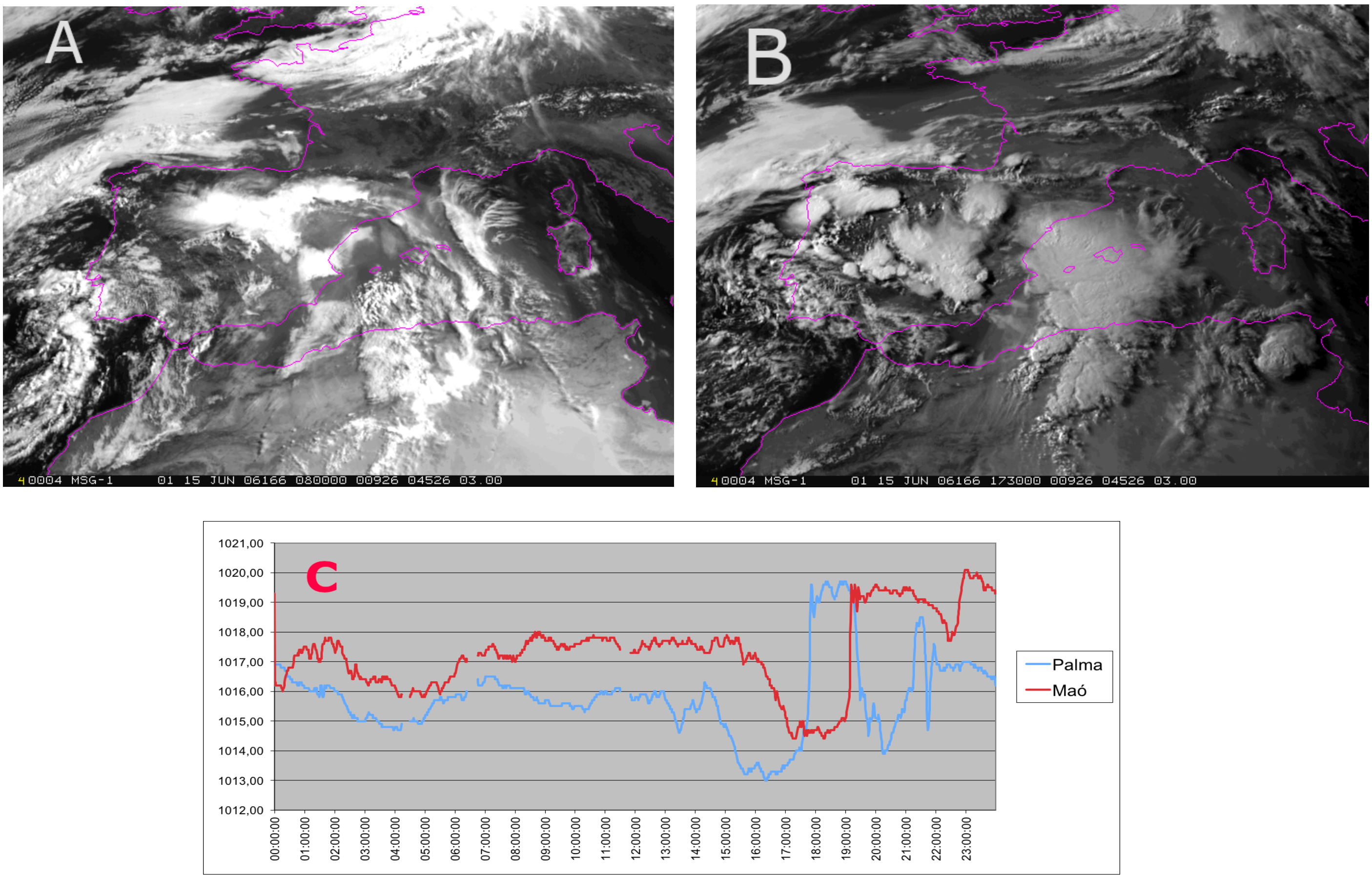

**Fig. 8**. *Satellite imagery and pressure records on 15 June 2006: a) Meteosat visible image at 08:00 GMT, b) Meteosat visible image at 17:30 GMT, and c) high-resolution pressure records from AEMET at Palma and Maó for the whole day.*

Since the pioneering work of Fujita (1955), it has been well established that convective pressure jumps are usually accompanied by squalls, that is, by strong wind gusts, with a very short duration. The occurrence of squalls or sudden and short duration strong winds therefore provides an indirect way to classify the situation (and so the *rissaga* event) as a convective one. Note that in the two most damaging *rissaga* events, (1984 and 2006), squalls accompanied the pressure jumps and high sea-level oscillations.

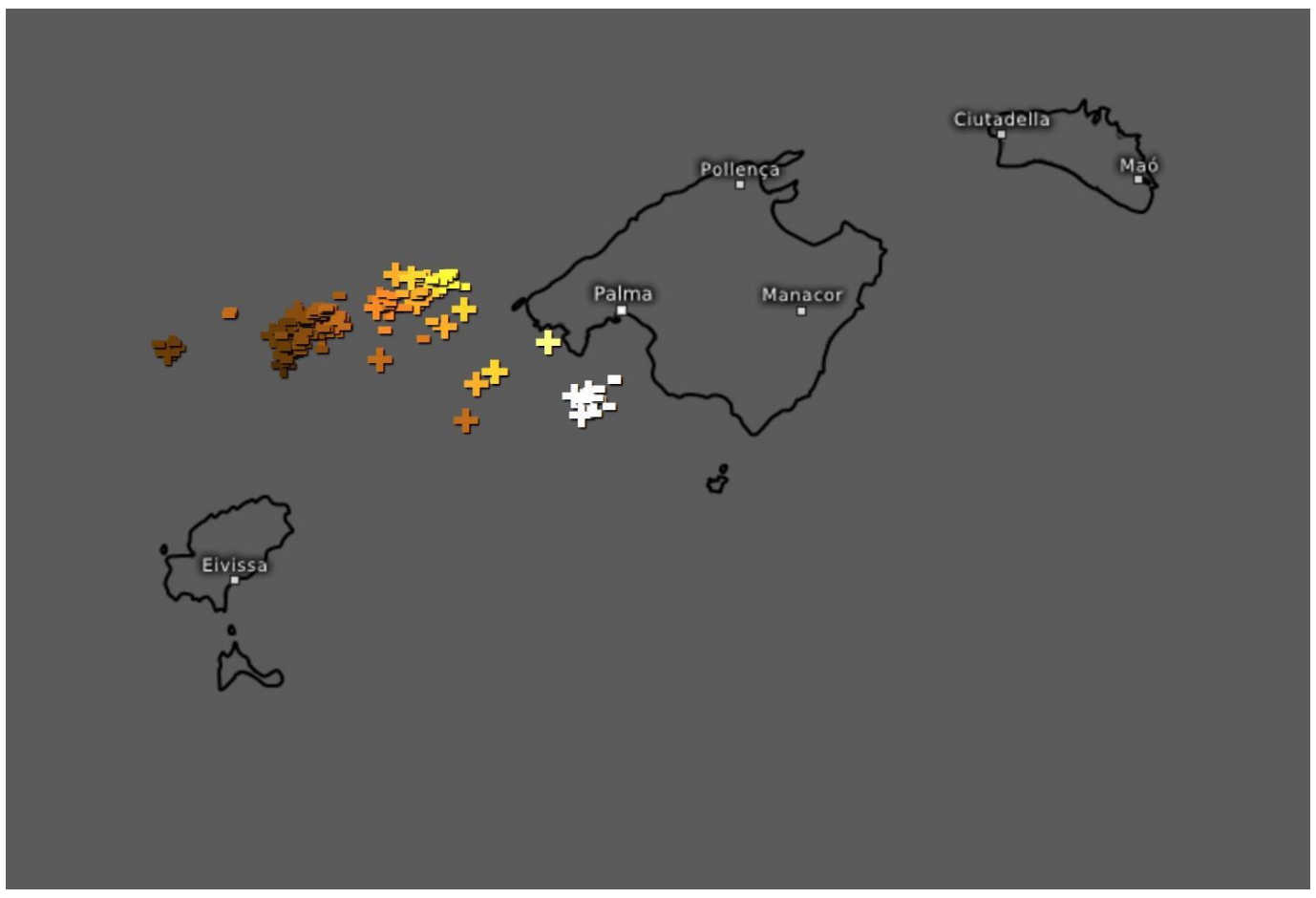

**Fig. 9**. *Lightning records on 12 July 2021 at 07 CEST, showing a convective system approaching Mallorca and Menorca. Image from Blitzortung.org.*

Another complementary method for assessing the presence of convective systems is the use of lightning maps. The detection of lightning provides unequivocal evidence of convection, and any associated *rissaga* event must therefore be classified as a convective event. **Fig. 9** illustrates such a case, corresponding to 12 July 2021.

### 2.4.3 Case classification

A case is considered as convective when convection is present in the area around Ciutadella for at least part of its duration, even if only briefly. However, it is often difficult to detect with certainty either the presence or the complete absence of convection. To address this uncertainty, the 191 events in our dataset were classified into the following 5 categories, depending on the degree of evidence of the presence of convection:

- A: High probability or clear evidence of a convective event
- B: Convective origin more likely than non-convective.
- C: Insufficient information to determine the type of event.
- D: Non-convective origin more likely than convective.
- E: High probability or clear evidence of a non-convective event.

## 2.5 Forecast performance of convective vs non-convective events

In this work, we further assess the relative predictability of *rissaga* events of convective origin compared with those of non-convective origin. Given that both UIB and SOCIB routinely provide objective, quantitative forecasts using the TRAM and BRIFS systems (Romero et al., 2019; Mourre et al., 2021), we retrieved the forecast amplitude of the *rissaga*. Note that several forecast amplitudes may be produced by each method during an event; we use the maximum value from each method and compare it with the maximum observed amplitude, both for convective and non-convective cases. For events without archived forecasts, hindcasts were performed when possible.

The TRAM method (Romero et al., 2019) is a hybrid atmospheric-oceanic modelling approach designed to capture, at low computational cost, the key physical processes responsible for many *rissaga* events. It consists of:

1. Atmospheric component: Genesis and north-eastward propagation of high-amplitude atmospheric gravity waves over the Balearic Islands. These waves are synthetically triggered and driven using a 2D non-hydrostatic, fully compressible model (a simplified version of the full TRAM model of Romero, 2023) in a dry-adiabatic atmosphere initialized with the Palma thermodynamic sounding. The horizontal resolution is 300 m while in the vertical it varies gradually from 20 m at sea level to 180 m in the uppermost computational layers.
2. Oceanic response in the Menorca channel (80 m depth): Long-wave amplification through Proudman-type resonance.
3. Shelf amplification: Doubling of wave amplitude through shoaling, as described by Green's law, when propagating from 80 m to 5 m depth.
4. Harbour resonance in Ciutadella: Further amplification within the 5 m deep inlet.

Steps (2) and (4) are modelled with 1D shallow-water equations (horizontal resolutions of 600 m and 12 m, respectively), while step (3) follows Green's law.

The TRAM scheme has demonstrated good skills in identifying *rissaga* situations and classifying events as weak, moderate or intense (Romero et al., 2019). Its main advantage is computational speed: the full procedure –from reading atmospheric sounding data to computation of the maximum sea-level oscillation– runs in ~5 minutes on a standard PC cluster. Operational results are generated twice daily, forced by GFS forecast soundings, and are publicly available at *http://meteo.uib.es/rissaga*. TRAM forecasts or hindcasts have been available since the beginning of the event record (1975).

The BRIFS system (Renault et al., 2011; Ličer et al., 2017; Mourre et al., 2021; *https://www.socib.es/es/que-hacemos/prediccion-del-oceano/modelo-brifs*) is based on fully realistic, 3D, high-resolution atmospheric and oceanic simulations. For the atmospheric component, the WRF model (Skamarock et al., 2008) is implemented in a two-way nested domain. The outer domain covers the western Mediterranean basin with a horizontal grid resolution of 20 km, while the inner domain reaches a horizontal grid resolution of 4 km, and it is centered on the Balearic Islands. Outputs from the inner domain, available at 1-minute temporal resolution , are used to force the *Regional Ocean Modeling System* (ROMS; Shchepetkin and McWilliams, 2005), configured to capture high-frequency sea-level variability. The ROMS configuration uses two nested grids: (i) a parent grid with 1 km grid resolution covering the shelves around Mallorca and Menorca, and (ii) an inner domain with 10 m grid resolution centred on the Ciutadella inlet. Further details on the modelling system can be found in Mourre et al., 2021. BRIFS has been producing daily *rissaga* forecasts since 2014. Additional BRIFS simulations have been carried out to represent the most relevant events included in our historical list before 2014, such as the event of June 15, 2006.

While BRIFS is considerably more computationally expensive than TRAM due to its fully realistic setup, a better representation of *rissaga* events could be expected, particularly under convective disturbances (TRAM is a dry-physics method). In general, it is found in both systems that realistic marine responses are achieved when the atmospheric pressure disturbances, either convective or non-convective, are well reproduced.

The *rissaga* amplitudes predicted by both systems are included in Appendix 1.

## 3. Some examples

### 3.1 Convective cases

The extraordinary and catastrophic cases of 21 June 1984 (Fig. 5) and 15 June 2006 (Fig. 8) have been studied in previous papers (Jansà, 1986; Jansà, 2007; Jansà and Ramis, 2021). Both are clear examples of convective cases.

A more complex case is the event of 2-3 July 1985, which led to a 3m sea level oscillation in Ciutadella, the third-highest amplitude in our listing. In the absence of pressure data in Ciutadella, the pressure register from Palma (Mallorca) is used to provide insights into the nature of the pressure variations. Rapid pressure oscillations are visible in Palma during several hours (not shown), but with a small magnitude (less than 1 hPa). If the pressure record in Palma accurately reflected the variations in Ciutadella, the resulting amplification factor would be unrealistically large—around 300.

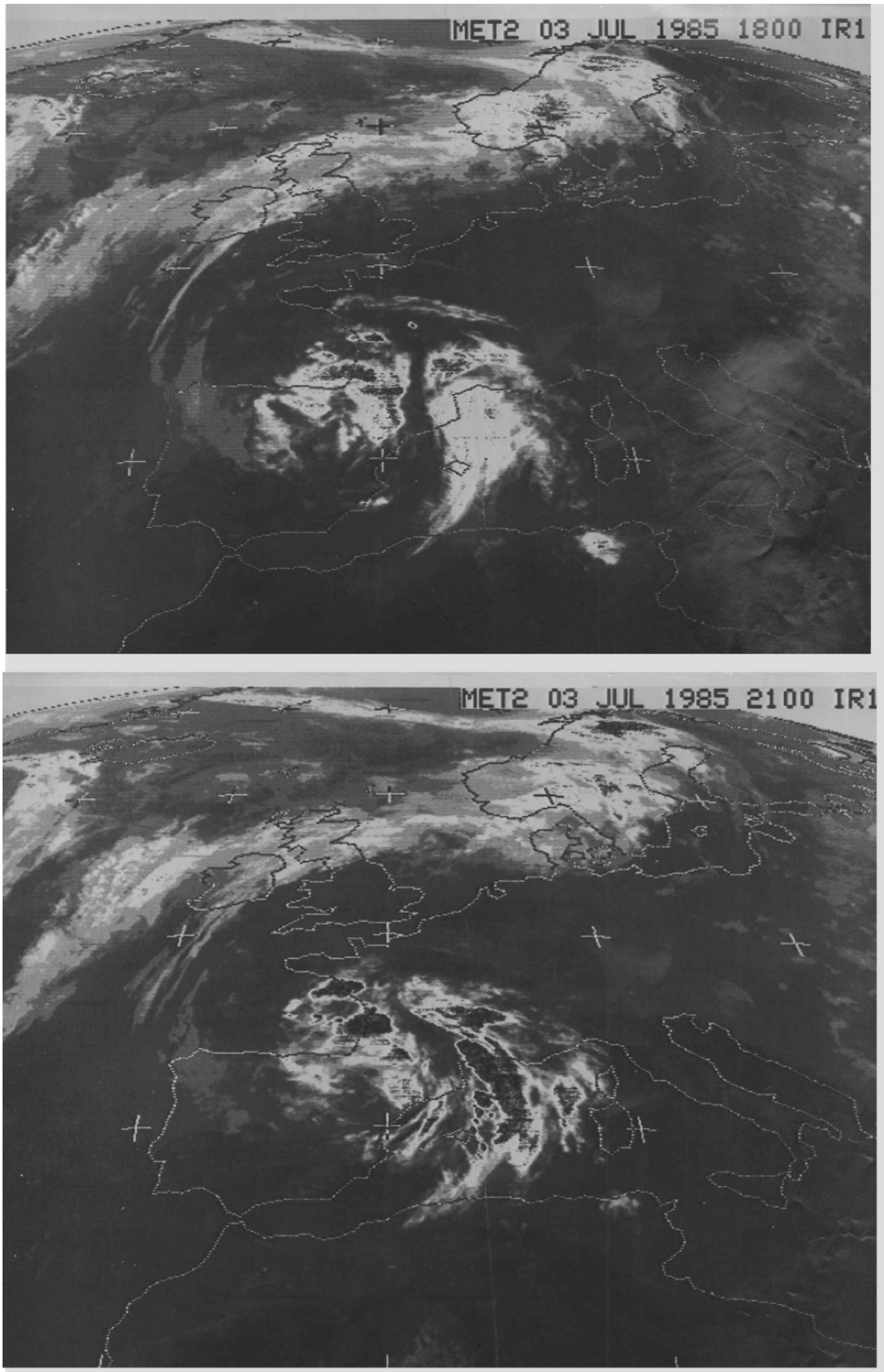

**Fig. 10**. *Meteosat IR imagery on 3 July 1985, at 18 UTC and at 21 UTC. The grayscale has been modified to enhance the visibility of high clouds. The modified scale progress from black to white (warm to cold).*

Looking at the IR satellite images (**Fig. 10**), we can interpret that significant changes in the cloudiness structure have occurred between 18 and 21 UTC of 3 July 1985 in the northern part of Mallorca island (but not in Palma) and to the north of the Balearics. At 18 UTC, a large stratiform comma cloud, with weak gravity waves embedded (not clearly visible), covers most of the region. At 21 UTC the cloudiness changed its structure with the occurrence of cold and well-defined cloud borders, which are representative of convective clouds. These cloudiness structures suggest that large, rapid pressure variations have probably occurred over the Menorca Channel.

Another interesting convective case is the event of 18-19 August 2014. The sea-level oscillation reached 147 cm, associated with an observed pressure variation of 2.6 hPa. This pressure change consisted in a sudden drop of pressure just followed by a surge (a "V" shaped perturbation). The estimated factor of amplification was 57, which is relatively moderate. This case was studied by Jansà (2014) and was mentioned by Ličer et al., (2017) as an example of efficient Proudman resonance during the propagation of the pressure perturbation over the Menorca Channel. A large sea-level oscillation was coincident in time with the pressure perturbation, also corresponding with the arrival at Ciutadella of an isolated small cloud nucleus, which is considered to be convective due to the shape of the atmospheric pressure signal.

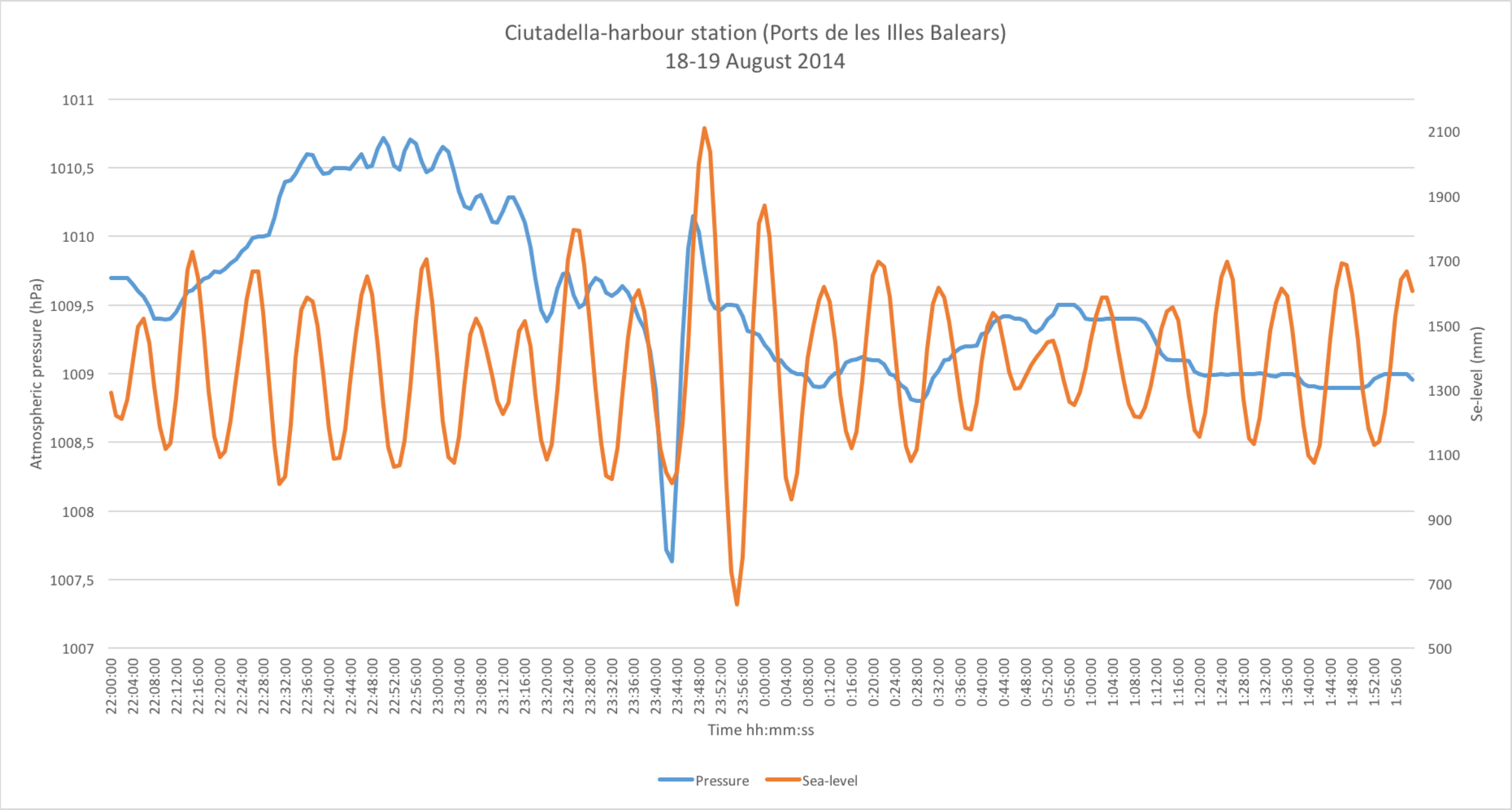

**Fig. 11**. *Atmospheric pressure (blue line) and sea-level (orange line) registers at the station of Port de Ciutadella (belonging to the Ports IB Agency) on 18-19 August 2014.*

A moderate *rissaga* occurred on 22 January 2021 (Jansà et al., 2025) during the winter season, which is rarely associated with significant meteotsunami events. This event was associated with a squall line which crossed most of the Balearic region, accompanied by a pressure surge of up to 4 hPa (2,9 hPa in Ciutadella). While sea-level oscillations were observed in several ports or inlets of the Balearics, with an amplitude around 25-30 cm, these oscillations reached 60 cm in Ciutadella. This clear example of convective case is associated with a relatively low amplification factor (21).

### 3.2 Non-convective cases

In cases defined as probably non-convective events, the pressure perturbations generally do not exhibit any significant singular changes, showing instead relatively regular variations. One example is the case of 2 July 1981, previously illustrated in **Fig. 4**.

Another remarkable case with quite regular pressure variations is the case of 23-24 June 2007. The registers of pressure and sea-level in Ciutadella depict the rapid and quite regular oscillations with an amplitude around 1.5 hPa occurring during a 12-hour period between 23 June 21:00 (local time) and 23 June 09:00. During this period, the sea level in Ciutadella oscillates with an amplitude around 100 cm, including peaks of up to 150 cm, corresponding to an amplification factor around 100. This large amplification suggests that both the Proudman resonance over the shelf and the harbor resonance in the inlet, which are linked to the relative regularity, propagation speed and period of the pressure oscillations, have been important in this case.

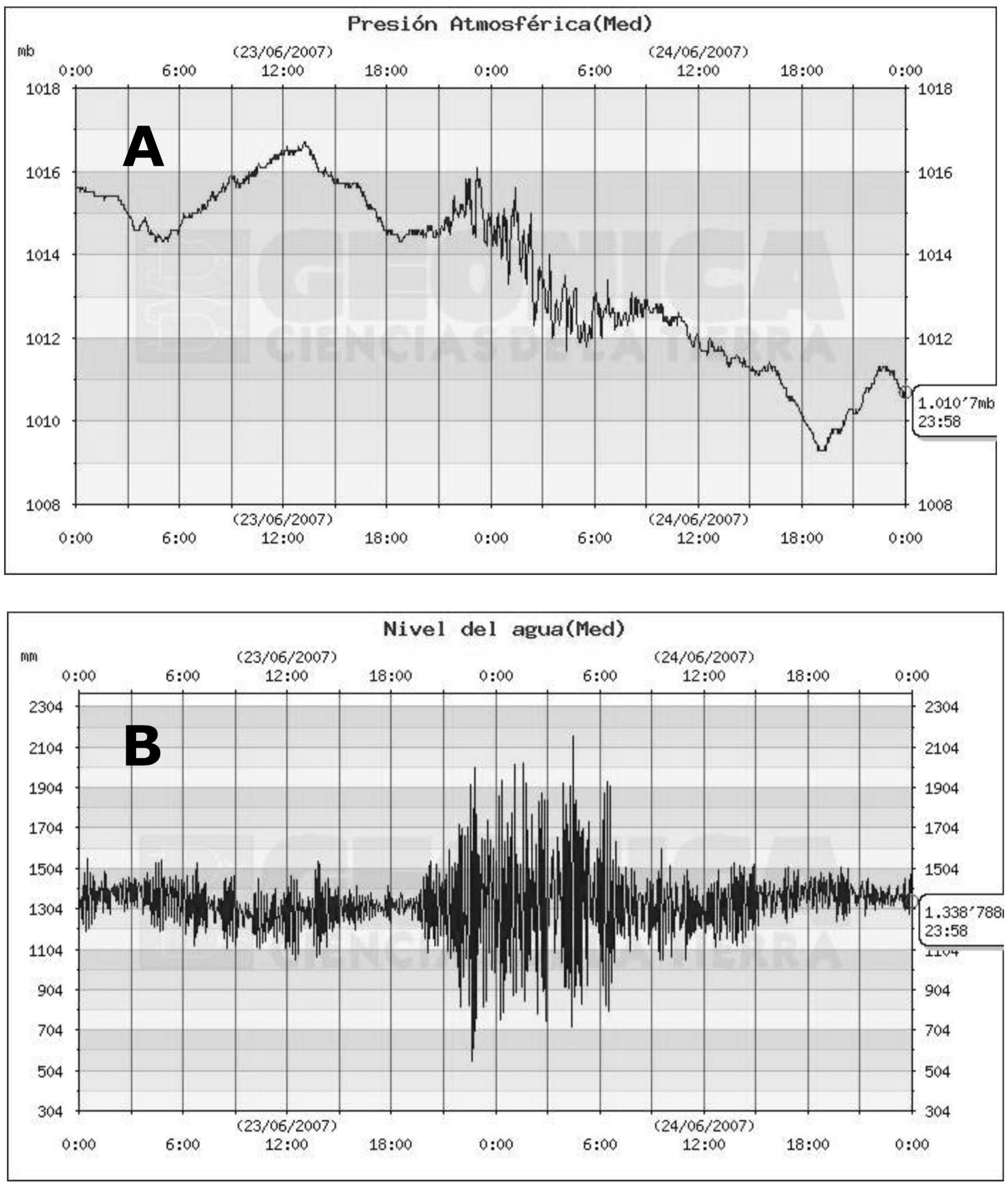

**Fig. 12**. *Records of pressure (a) and sea-level (a) in Ciutadella, obtained from Ports IB sensors.*

## 4. Results

In this section, we present and discuss results related to amplitude, pressure perturbations, amplification factors, and prediction quality for both convective and non-convective events.

According to the event classification described in Section 2.4.3, the full set of 191 events has been grouped as follows: 21 cases are Type A (clearly convective), 42 are Type B (probably convective), 42 are Type C (uncertain classification), 53 are likely non-convective, and 33 are clearly non-convective. This categorization results in 63 events with a moderate to high probability of being convective and 86 that are likely non-convective. Although non-convective events are more numerous, convective cases remain highly relevant.

### 4.1 Amplitude

**Fig. 13** shows the amplitudes of the full series of events. The low-pass filtered line reveals an abrupt change around 2007–2008. In 2007, a permanent meteo-oceanographic station was installed in Ciutadella, marking the beginning of systematic, high-resolution digital recordings of sea level and atmospheric pressure.

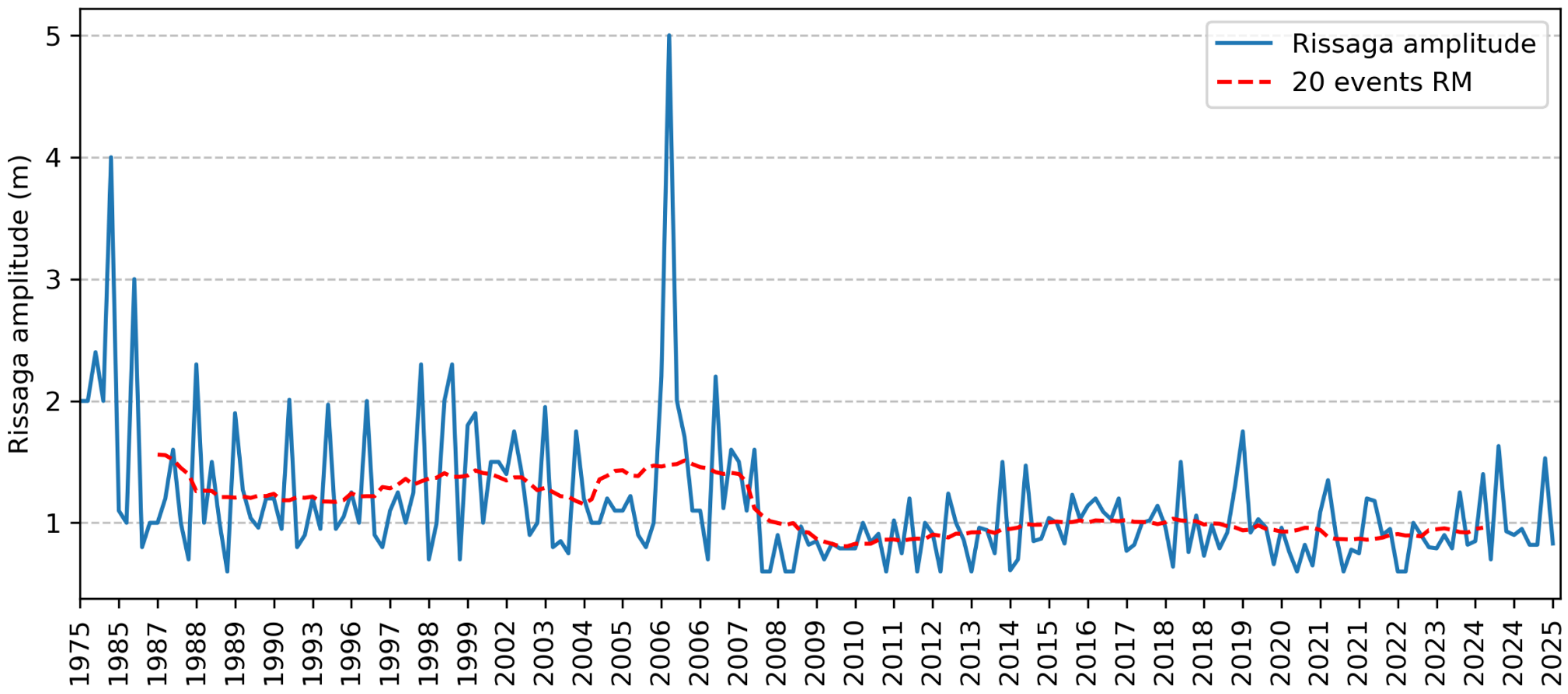

**Fig. 13.** *Chronological series of rissaga amplitudes from 1975 to 2025 (blue line). The dashed red line represents a low-pass filter obtained using a running average over 20 events.*

The average amplitude for the events before the steep change (1975–2007) is 138 cm, whereas the average amplitude after (2008–2025) is 94 cm. Although this difference is considerable, the abrupt transition suggests it is not driven by long-term environmental changes. As mentioned in Section 2.2, there are at least two other plausible explanations for this discrepancy.

First, aside from the digital observations collected during field campaigns, most records prior to 2007–2008 were based on subjective estimates. As a result, less significant *rissagas*, particularly those occurring at night, may have been unrecorded, while the largest events were almost always documented. Second, those subjective estimates were typically made near the end of the harbor (i.e., the point farthest from the open sea), where sea-level oscillations are usually strongest. In contrast, the instrumental records are taken from an intermediate location, where the amplitude of sea-level variations is lower than at the harbor's end.

### 4.2 Relationship between type of event and magnitude

**Table 1** shows the frequency of the *rissaga* events as a function of the *rissaga* amplitude and type (i.e., convective or not).

**Table 1**. *Frequency of rissaga events in Ciutadella, as a function of amplitude and type*
*A:* High probability or clear evidence of a convective event
*B:* Convective origin more likely than non-convective.
*C:* Insufficient information to determine the type of event.
*D:* Non-convective origin more likely than convective.
*E:* High probability or clear evidence of a non-convective event.

| *Amplitude* | *Total* | *A* | *B* | *A+B* | *C* | *D* | *E* | *D+E* |
|---|---|---|---|---|---|---|---|---|
| *>60* | *191* | *21* | *42* | *63* | *42* | *53* | *33* | *86* |
| *>80* | *152* | *16* | *37* | *53* | *29* | *44* | *26* | *70* |
| *>100* | *103* | *13* | *21* | *34* | *23* | *29* | *17* | *46* |
| >120 | 61 | 10 | 8 | 18 | 13 | 19 | 11 | 30 |
| >140 | 40 | 9 | 6 | 15 | 6 | 12 | 7 | 19 |
| >160 | 29 | 7 | 5 | 12 | 5 | 8 | 4 | 12 |
| >180 | 21 | 5 | 4 | 9 | 2 | 6 | 4 | 10 |
| >200 | 16 | 5 | 3 | 8 | 2 | 4 | 2 | 6 |
| >220 | 9 | 4 | 2 | 6 | 0 | 3 | 0 | 3 |
| >240 | 4 | 3 | 0 | 3 | 0 | 1 | 0 | 1 |
| >260 | 3 | 3 | 0 | 3 | 0 | 0 | 0 | 0 |
| >280 | 3 | 3 | 0 | 3 | 0 | 0 | 0 | 0 |
| >300 | 3 | 3 | 0 | 3 | 0 | 0 | 0 | 0 |

It can be observed that the total number of events with a higher probability of convective origin (categories A + B) is lower than the number of non-convective events (categories D + E). Interestingly, the proportion of A + B events relative to D + E increases with rising *rissaga* amplitude.

Although the overall ratio (A + B) / (D + E) is approximately 0.73 when all cases are considered—and this ratio holds for events with amplitudes greater than 100 cm—it slightly increases to 0.79 for amplitudes above 140 cm, and continues to rise with event magnitude (0.9 for events > 180 cm; 2.0 for events > 220 cm). However, this increase is not linear. Importantly, extreme amplitude events correspond exclusively to convective cases.

Another useful metric to consider is the *amplification factor*, defined as the ratio between the maximum atmospheric pressure variation and the amplitude of the corresponding *rissaga*. For example, an amplification factor of 100 means that a pressure perturbation of just 1 hPa would result in a sea-level oscillation of 1 meter. **Table 2** presents the amplification factor as a function of event type (convective or non-convective). On average, the amplification factor is 71.

This value could only be computed when both *rissaga* amplitude and pressure data were available. As a result, Table 2 includes only 115 events for which sufficient data were available, out of the total 191. Notably, in all cases, the amplification factor exceeds 15.

When the amplification factor is relatively moderate (up to around 30), the number of convective and non-convective cases is quite similar. However, the highest amplification factors tend to correspond more frequently to non-convective events, although the difference is not particularly large. For example, amplification values greater than 50 are observed in 32 non-convective cases, compared to only 24 convective ones.

**Table 2**. *Factor of amplification (FA) of rissaga events in Ciutadella, as a function of their type*
*A:* High probability or clear evidence of a convective event
*B:* Convective origin more likely than non-convective.
*C:* Insufficient information to determine the type of event.
*D:* Non-convective origin more likely than convective.
*E:* High probability or clear evidence of a non-convective event.

| FA (>=) | *Tot* | *A* | *B* | *A+B* | *C* | *D* | *E* | *D+E* |
|---|---|---|---|---|---|---|---|---|
| 15 | 115 | 21 | 32 | 53 | 13 | 27 | 22 | 49 |
| 30 | 102 | 19 | 30 | 49 | 9 | 23 | 21 | 44 |
| 50 | 61 | 12 | 12 | 24 | 5 | 16 | 16 | 32 |
| 100 | 17 | 3 | 4 | 7 | 1 | 5 | 4 | 9 |
| 200 | 5 | 1 | 1 | 2 | 1 | 2 | 0 | 2 |
| 300 | 0 | 0 | 0 | 0 | 0 | 0 | 0 | 0 |

**Table 3** explores the possible relationship between the magnitude of pressure variations and the type of *rissaga* event. The most significant pressure changes tend to be associated with convective cases (A+B), which show higher values than non-convective cases (D+E). Specifically, 39 A+B cases recorded pressure changes equal to or greater than 2 hPa, compared to only 18 such cases in the D+E category. Pressure variations of 4 hPa or more were observed in just one case from each group. Interestingly, only one event—classified as a convective origin *rissaga*—recorded a pressure variation equal to or exceeding 6 hPa.

**Table 3**. *Relationship between pressure variation ($\Delta P$) and type of rissaga*
*A:* High probability or clear evidence of a convective event
*B:* Convective origin more likely than non-convective.
*C:* Insufficient information to determine the type of event.
*D:* Non-convective origin more likely than convective.
*E:* High probability or clear evidence of a non-convective event.

| $\Delta P$ | A | B | A+B | C | D | E | D+E | TOT |
|---|---|---|---|---|---|---|---|---|
| >=2 | 15 | 22 | 37 | 7 | 13 | 5 | 18 | 62 |
| >=4 | 1 | 0 | 1 | 1 | 1 | 0 | 1 | 3 |
| >=6 | 1 | 0 | 1 | 0 | 0 | 0 | 0 | 1 |

In summary, convective *rissaga* events can reach higher final amplitudes than non-convective ones, with initial pressure perturbations larger than in non-convective cases and a smaller amplification factor.

### 4.3. Predictability analysis

Given the significance of *rissaga* events and their potential impact on infrastructure in Ciutadella harbor, as well as the risks they pose to human safety, it is essential to assess the predictability of these natural phenomena in the Balearic Islands. To this end, the forecast amplitudes from the TRAM and BRIFS methods have been compared with the observed *rissaga* amplitudes. **Table 4** presents Pearson's correlation coefficients between each forecasting system and observations. Interestingly, BRIFS gives more accurate forecasts for convective cases compared to non-convective cases. The opposite occurs with the TRAM method: their forecasts are better for non-convective cases.

**Table 4**. *Pearson correlation coefficient between forecasts and observed total cases, convective cases (A+B) and non-convective cases (D+E)*
*A:* High probability or clear evidence of a convective event
*B:* Convective origin more likely than non-convective.
*C:* Insufficient information to determine the type of event.
*D:* Non-convective origin more likely than convective.
*E:* High probability or clear evidence of a non-convective event.

| *Method* | *Total* | *A+B* | *D+E* |
|---|---|---|---|
| TRAM | 0,29 | 0,28 | **0,48** |
| BRIFS | 0,53 | **0,65** | 0,35 |

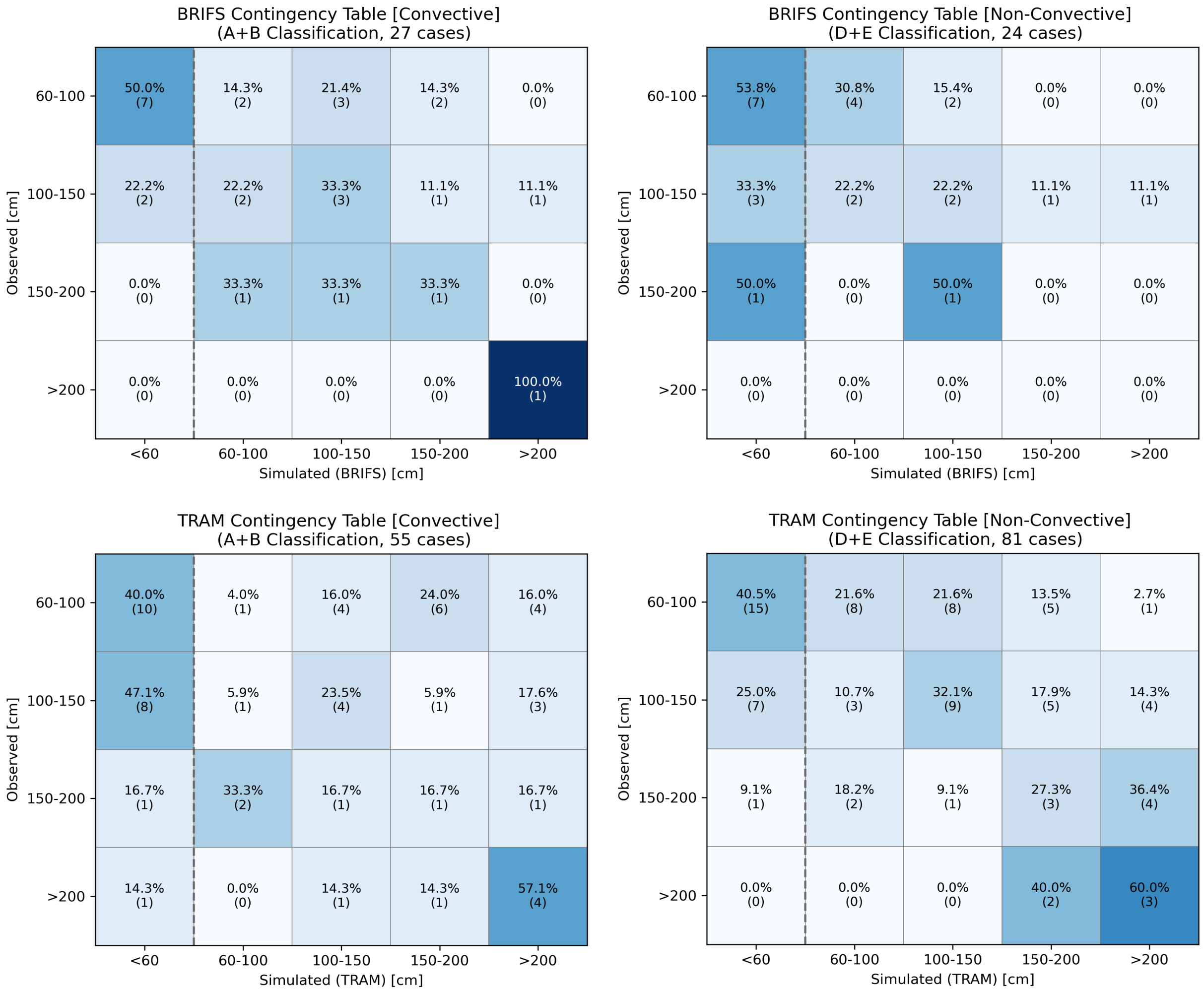

**Fig. 14.** *Contingency tables for predicted vs observed rissaga amplitudes, for convective (A+B) and non-convective (D+E) events, provided by the BRIFS and TRAM methods.*

Further assessment of predictive skill was carried out by constructing contingency tables for forecast vs observed *rissaga* height, stratified by four amplitude categories (60–100, 100–150, 150–200, and >200 cm), considering separately the clear and probable convective cases (A+B) and the clear and probable non-convective (D+E) events (Fig. 15). The results highlight both the limitations and the contrasting behavior of the two predictive systems (BRIFS and TRAM). Overall, both systems exhibit a substantial scatter of values away from the descending diagonal, underscoring the inherent difficulty of providing accurate estimates of *rissaga* intensity.

Despite this general uncertainty, some systematic tendencies can be identified. BRIFS performs comparatively better under convective conditions, showing a higher concentration of correct classifications. This suggests that the system is able to capture, at least to some extent, the convective forcing mechanisms responsible for *rissaga* generation. Such behavior is consistent with the expectation that convective processes are explicitly represented in the atmospheric models that drive the BRIFS system. By contrast, its performance under non-convective cases appears weaker, with predictions showing greater dispersion and less agreement with the observed categories.

The TRAM system exhibits the opposite behavior, performing more favorably during non-convective events, where simulated and observed values show a clearer correspondence. In convective situations, however, the predictions display a markedly higher degree of dispersion, reflecting the inherent limitations of dry-adiabatic simulations in reproducing mesoscale disturbances associated with moist convection. This outcome is consistent with the design of the TRAM system, which relies primarily on synoptic-scale patterns and dry dynamics and therefore does not explicitly incorporate moist convective forcing as an isolated mechanism in the development of rissaga events.

Another noteworthy aspect is the ability of both systems to reproduce extreme events, particularly those exceeding 200 cm. Although the number of cases in this category is small and the result cannot be considered statistically robust, it nonetheless represents a highly relevant finding, since such extreme episodes almost invariably lead to severe disruptions and substantial economic losses.

In summary, the results indicate that both predictive systems provide a valuable qualitative assessment of *rissaga* risk, though not precise quantitative forecasts. Their respective strengths are clearly differentiated: BRIFS shows greater relative skill in convective situations, whereas TRAM proves more effective in non-convective cases. Taken together, these findings point to the potential benefit of a complementary application of both systems, with the choice of predictive tool conditioned by the prevailing atmospheric forcing.

## 5. Conclusions

From 1975 to 2025, a total of 191 *rissaga* events with amplitudes ≥60 cm were documented in Ciutadella harbor (Menorca) corresponding to an average of about four events per year. We examine these cases to quantify the frequency and characteristics according to the origin of the pressure disturbance that generated the *rissaga.* Specifically, two different types are identified: (a) convective *rissaga*: those linked to rapid pressure variations mainly associated with convective systems, and (b) non-convective *rissaga*: those linked to pressure variations mainly associated with internal gravity waves without convective involvement.

To classify the *risssaga* events as convective or non-convective, we relied on three elements: the shape of the pressure record, the presence of convective clouds in satellite imagery, and the occurrence of lightning. Applying these criteria, we found that convective events are slightly less frequent than non-convective ones, but the difference is not large. Of the 191 cases, 63 (33%) were clearly or likely convective, 86 (45%) were clearly or likely non-convective, and 42 (22%) could not be classified. Aside from issues related to data unavailability, the uncertainty of the classification results reflects the fact that some steps in the procedure involve a certain degree of subjectivity.

Regarding amplitude, the first point to note is that the average amplitude value during the earlier period (1975-2007) is clearly higher than in more recent years (2007-2025). In 2007, permanent instrumentation was installed in Ciutadella, allowing more detailed records of sea-level and atmospheric pressure. The differences between the two periods can be explained by the used methodology: non-instrumental observations were more likely to miss some events, and measurements also depended on the exact location within the port.

Leaving these differences aside, Table 1 indicates that the ratio of the amplitude of convective to non-convective events increases with the *rissaga* amplitude, although not linearly. The ratio is 0,7 when all cases ≥60 cm are considered, and remains 0,7 for amplitudes ≥100 cm. For amplitudes ≥ 140 cm the ratio rises to 0,8, and for amplitudes ≥180 cm it is 0,9. If the amplitude exceeds 220 cm, the ratio is 2,0. Extreme amplitude events are exclusively associated with convective cases.

Regarding the amplification factor, defined as the *rissaga* height (cm) divided by the atmospheric pressure change (hPa), the average value is 71. For moderate factors, up to about 30, the number of convective and non-convective cases is similar. Higher amplification factors occur more often in non-convective events, although the difference is not large: values above 50 were found in 32 non-convective cases and in 24 convective cases.

The relationship with pressure variation is clearer. Convective events generally exhibit stronger pressure perturbations than their non-convective counterparts. In 39 convective cases, the variation was ≥2 hPa, compared with only 18 non-convective cases. Pressure jumps ≥4 hPa were recorded in one convective and one non-convective event, while an extreme variation ≥6 hPa was observed only in a convective case.

Regarding predictability, while the convective cases are found to be more predictable when a full-realistic atmosphere and ocean combined modelling system is used (BRIFS), the reduced-physics TRAM method is found to provide better predictions for non-convective events. These results highlight the complementarity of the two kinds of prediction systems to properly represent the different types of triggering mechanisms.

***Acknowledgements.***
D. S. Carrió, R. Romero and V. Homar acknowledge the Spanish HYDROMED grant (PID2023-146625OB-I00) funded by MICIU/AEI/10.13039/501100011033 under ERDF/EU.

**Conflict of Interest**
Not applicable

## Appendix 1: Historical catalogue of *rissaga* events in Ciutadella (1975-2025).

*OBSERVED = maximum observed rissaga amplitude (cm)*
$\Delta P$ *= magnitude of the maximum atmospheric pressure changes (hPa)*
*Factor: total factor of amplification, from the inverse barometer effect to the final rissaga amplitude*
*A: high probability or clear evidence of a convective event*
*B: convective origin more likely than non-convective*
*C: insufficient information to determine the type of event*
*D: non-convective origin more likely than convective*
*E: high probability or clear evidence of a non-convective event*
*TRAM: maximum rissaga amplitude (cm) predicted by TRAM.*
*BRIFS: maximum rissaga amplitude (cm) predicted by BRIFS.*
*NA: Not Available*

| Year | Month | Day(s) | OBSERVED Amplitude | *ΔP* | Factor | A | B | C | D | E | TRAM Amplitude | BRIFS Amplitude |
|---|---|---|---|---|---|---|---|---|---|---|---|---|
| 1975 | 9 | 16 | 200 | 1,5 | 133 | | | | | X | 245,1 | NA |
| 1981 | 7 | 2 | 200 | 1,5 | 133 | | | | | X | 235,9 | NA |
| 1982 | 6 | 18-19 | 240 | 2,5 | 96 | | | | X | | 173,1 | NA |
| 1982 | 7 | 29-02 | 200 | 2 | 100 | | | | X | | 344,3 | NA |
| 1984 | 6 | 21 | 400 | 3 | 133 | X | | | | | 358,3 | NA |
| 1985 | 6 | 14 | 110 | NA | NA | | | | X | | 205,2 | NA |
| 1985 | 6 | 19-20 | 100 | NA | NA | | | X | | | 91,2 | NA |
| 1985 | 7 | 2-3 | 300 | 1 | 300 | X | | | | | 204,8 | NA |
| 1985 | 7 | 31 | 80 | NA | NA | | | | | X | 43,7 | NA |
| 1985 | 8 | 5 | 100 | NA | NA | | | | X | | 38,9 | NA |
| 1987 | 6 | 15 | 100 | NA | NA | | | | | X | 123,1 | NA |
| 1987 | 6 | 23 | 120 | NA | NA | | | X | | | 24,5 | NA |
| 1987 | 7 | 8 | 160 | NA | NA | | | X | | | 33,8 | NA |
| 1987 | 7 | 15 | 100 | 2 | 50 | | X | | | | 22,6 | NA |
| 1987 | 8 | 9 | 70 | NA | NA | | | X | | | 141,7 | NA |
| 1988 | 5 | 6-7 | 230 | 2,5 | 92 | | X | | | | NA | NA |
| 1988 | 6 | 29 | 100 | 1 | 100 | | | | | X | NA | NA |
| 1988 | 7 | 7-8 | 150 | 1 | 150 | | | | | X | 163,6 | NA |
| 1989 | 5 | 11-12 | 100 | 2 | 50 | | | | X | | 178,7 | NA |
| 1989 | 5 | 22 | 60 | NA | NA | | | | X | | 45,1 | NA |
| 1989 | 7 | 5-8 | 190 | 3,5 | 54 | | X | | | | 96,9 | NA |
| 1989 | 8 | 5 | 127 | NA | NA | | | | X | | 39 | NA |
| 1989 | 8 | 10 | 104 | NA | NA | | | | X | | 15,7 | NA |
| 1989 | 9 | 10 | 96 | NA | NA | | | | | X | 103,2 | NA |
| 1990 | 5 | 23 | 120 | NA | NA | | X | | | | 37,1 | NA |
| 1990 | 6 | 11 | 120 | NA | NA | | | | X | | 141,6 | NA |
| 1990 | 9 | 25-26 | 95 | 2,5 | 38 | | | | X | | 219,1 | NA |
| 1992 | 6 | 21 | 200 | 3 | 67 | | X | | | | 108 | NA |
| 1992 | 8 | 29 | 80 | 1,5 | 53 | | | | X | | 36,8 | NA |
| 1993 | 7 | 3 | 90 | NA | NA | | | | | X | 107,9 | NA |
| 1993 | 7 | 10 | 120 | NA | NA | | | | | X | 112,9 | NA |
| 1993 | 9 | 23 | 95 | NA | NA | | X | | | | 10,5 | NA |
| 1994 | 6 | 2 | 197 | NA | NA | | | | | X | 83,9 | NA |
| 1994 | 6 | 19 | 95 | NA | NA | | | | | X | NA | NA |
| 1994 | 6 | 26 | 105 | NA | NA | | | X | | | 129 | NA |
| 1996 | 5 | 25 | 125 | NA | NA | | | | X | | 47 | NA |
| 1996 | 7 | 29 | 100 | NA | NA | | | X | | | 113,6 | NA |
| 1996 | 7 | 31 | 200 | NA | NA | | | X | | | 8,5 | NA |
| 1996 | 8 | 15 | 90 | NA | NA | | X | | | | 93,8 | NA |
| 1996 | 9 | 11 | 80 | NA | NA | | X | | | | 23,2 | NA |
| 1997 | 6 | 7-11 | 110 | 2 | 55 | | | | | X | 150,6 | NA |
| 1997 | 7 | 1-4 | 125 | NA | NA | | | X | | | 179 | NA |
| 1997 | 7 | 14 | 100 | NA | NA | | | | X | | 118,8 | NA |
| 1997 | 7 | 23-25 | 125 | 3,5 | 36 | | | X | | | 215,3 | NA |
| 1997 | 7 | 31 | 230 | NA | NA | | | | X | | NA | NA |
| 1998 | 7 | 7 | 70 | NA | NA | | | | X | | 153,2 | NA |
| 1998 | 7 | 14 | 100 | NA | NA | | | | X | | 111,8 | NA |
| 1998 | 7 | 30 | 200 | NA | NA | | | X | | | 27,9 | NA |
| 1998 | 8 | 3 | 230 | NA | NA | | | | X | | 171,6 | NA |
| 1999 | 6 | 7 | 70 | NA | NA | | | | X | | 74,9 | NA |
| 1999 | 8 | 3 | 180 | NA | NA | | | | X | | 35,7 | NA |
| 1999 | 8 | 9-10 | 190 | NA | NA | | | | | X | 97 | NA |
| 2001 | 5 | 17 | 100 | NA | NA | | | | | X | 103,1 | NA |
| 2001 | 6 | 10 | 150 | NA | NA | | | | X | | 322,7 | NA |
| 2002 | 6 | 4-5 | 150 | NA | NA | | | | | X | 167,1 | NA |
| 2002 | 7 | 22 | 140 | NA | NA | | | X | | | 224,7 | NA |
| 2002 | 8 | 2 | 175 | NA | NA | | | X | | | 86,4 | NA |
| 2002 | 8 | 5 | 140 | NA | NA | | | | X | | 102,5 | NA |
| 2003 | 4 | 30 | 90 | NA | NA | | | X | | | 158,6 | NA |
| 2003 | 7 | 5 | 100 | NA | NA | | | X | | | 68,6 | NA |
| 2003 | 8 | 28 | 195 | NA | NA | | | | X | | 242,9 | NA |
| 2003 | 9 | 4 | 80 | NA | NA | | | X | | | 109,7 | NA |
| 2004 | 2 | 21 | 85 | NA | NA | | | X | | | 77,2 | NA |
| 2004 | 4 | 3 | 75 | NA | NA | | | X | | | 17,6 | NA |
| 2004 | 7 | 7-8 | 175 | 1,5 | 117 | | X | | | | 160,9 | NA |
| 2004 | 8 | 2 | 120 | 1,5 | 80 | | | | | X | 53,7 | NA |
| 2005 | 5 | 12 | 100 | NA | NA | | X | | | | 116,3 | NA |
| 2005 | 7 | 4 | 100 | NA | NA | | X | | | | 25,6 | NA |
| 2005 | 7 | 18 | 120 | NA | NA | | | | X | | 305 | NA |
| 2005 | 7 | 29 | 110 | NA | NA | | | X | | | 171,3 | NA |
| 2005 | 8 | 10-11 | 110 | NA | NA | | X | | | | 83,6 | NA |
| 2005 | 8 | 18 | 122 | NA | NA | | | X | | | 149 | NA |
| 2006 | 4 | 10 | 90 | NA | NA | | X | | | | 193,9 | NA |
| 2006 | 5 | 19 | 80 | NA | NA | | | | X | | 135,4 | NA |
| 2006 | 5 | 23-25 | 220 | NA | NA | | X | | | | 262,1 | NA |
| 2006 | 6 | 15 | 500 | 7 | 71 | X | | | | | 42,3 | 310 |
| 2006 | 6 | 17-19 | 200 | 3,5 | 57 | X | | | | | 234,1 | NA |
| 2006 | 6 | 27-28 | 170 | NA | NA | | | X | | | 87 | NA |
| 2006 | 8 | 18 | 110 | NA | NA | | | | X | | 149,8 | NA |
| 2006 | 8 | 20 | 110 | NA | NA | | | X | | | 147,3 | NA |
| 2006 | 8 | 24 | 70 | NA | NA | | | X | | | 52,6 | NA |
| 2007 | 5 | 22-23 | 100 | 2,5 | 40 | | X | | | | 37,8 | NA |
| 2007 | 6 | 17-18 | 112 | 2 | 56 | | | | X | | 115,1 | NA |
| 2007 | 6 | 22 | 160 | NA | NA | | | | X | | 174,2 | NA |
| 2007 | 6 | 23-24 | 150 | 1,5 | 100 | | | | | X | 264,7 | NA |
| 2007 | 7 | 22 | 110 | NA | NA | | | X | | | 208,2 | NA |
| 2007 | 10 | 4 | 160 | 2,5 | 64 | X | | | | | 138,2 | NA |
| 2008 | 5 | 24-27 | 220 | 3 | 73 | X | | | | | 187,3 | NA |
| 2008 | 5 | 31-1 | 60 | 0,5 | 120 | | | | X | | 59,1 | NA |
| 2008 | 7 | 14 | 90 | 2 | 45 | | X | | | | 112,2 | NA |
| 2009 | 4 | 24-25 | 60 | 2 | 30 | | | | X | | 57,9 | NA |
| 2009 | 6 | 25 | 60 | 3,5 | 17 | | | | X | | 81,3 | NA |
| 2009 | 7 | 8 | 97 | 1 | 97 | | | | | X | 51,4 | NA |
| 2009 | 7 | 14 | 82 | 2,5 | 33 | | | | X | | 122,1 | NA |
| 2009 | 7 | 27 | 85 | 2 | 43 | | | | | X | 36,5 | NA |
| 2010 | 2 | 27 | 60 | 0,5 | 120 | | | | | X | 183,4 | NA |
| 2010 | 6 | 9 | 70 | 1,5 | 47 | | | | | X | 60,5 | NA |
| 2010 | 6 | 14 | 83 | 1,5 | 55 | | | | X | | 71,1 | NA |
| 2010 | 7 | 24 | 79 | NA | NA | | | X | | | 128,8 | NA |
| 2010 | 7 | 27 | 79 | NA | NA | | | X | | | 7,3 | NA |
| 2010 | 8 | 10 | 100 | NA | NA | | | X | | | 6,9 | NA |
| 2010 | 9 | 18 | 84 | NA | NA | | | X | | | 75,3 | NA |
| 2011 | 7 | 4 | 79 | 2 | 40 | | X | | | | 114 | NA |
| 2011 | 7 | 12-13 | 60 | 1,5 | 40 | | X | | | | 151,4 | NA |
| 2011 | 7 | 22 | 102 | 3 | 34 | | X | | | | 203,4 | NA |
| 2012 | 4 | 10 | 91 | 2,5 | 36 | | | | | X | 52,6 | NA |
| 2012 | 4 | 28 | 75 | 1 | 75 | | | | | X | 158,7 | NA |
| 2012 | 6 | 7 | 60 | 1 | 60 | | | | | X | 90 | NA |
| 2012 | 6 | 20 | 100 | 3 | 33 | | X | | | | 39 | NA |
| 2012 | 7 | 12 | 91 | 2 | 46 | X | | | | | 177,1 | NA |
| 2012 | 7 | 14 | 60 | 2,5 | 24 | | | | X | | NA | NA |
| 2012 | 7 | 27 | 124 | NA | NA | | | X | | | 330,5 | NA |
| 2012 | 8 | 6 | 100 | 2 | 50 | | X | | | | 308,3 | NA |
| 2013 | 5 | 15 | 120 | NA | NA | | | X | | | 149,4 | NA |
| 2013 | 6 | 17-18 | 60 | 2 | 30 | | | | | X | 94,8 | NA |
| 2013 | 9 | 8 | 96 | NA | NA | | | | X | | 64,4 | NA |
| 2013 | 9 | 28 | 94 | NA | NA | | | | X | | 60,3 | NA |
| 2013 | 10 | 4 | 75 | 2 | 38 | X | | | | | 43,2 | NA |
| 2014 | 5 | 20 | 86 | 2,5 | 34 | X | | | | | 112,9 | NA |
| 2014 | 6 | 23 | 61 | 2,5 | 24 | X | | | | | 59,8 | NA |
| 2014 | 7 | 4 | 70 | NA | NA | | | X | | | 72,5 | NA |
| 2014 | 8 | 18-19 | 147 | 2,6 | 57 | X | | | | | 147,9 | 112 |
| 2014 | 9 | 16 | 85 | 1,5 | 57 | | X | | | | 58,9 | NA |
| 2014 | 9 | 19 | 87 | 2,5 | 35 | | X | | | | 134,8 | NA |
| 2015 | 4 | 22 | 150 | 1 | 150 | | X | | | | 68,1 | 66 |
| 2015 | 5 | 6 | 104 | 1,4 | 74 | | | X | | | 434,2 | 22 |
| 2015 | 7 | 29-1 | 130 | 3,3 | 39 | | | | | X | 173,4 | 207 |
| 2016 | 2 | 7 | 83 | 1,4 | 59 | | | | X | | 47,2 | 11 |
| 2016 | 4 | 1 | 123 | 1,1 | 112 | | | | X | | 43,9 | 51 |
| 2016 | 5 | 25 | 103 | 1,1 | 94 | | | | | X | 204,2 | 4 |
| 2016 | 5 | 28 | 114 | 3 | 38 | | X | | | | 113,2 | 61 |
| 2017 | 3 | 4 | 120 | 0,4 | 300 | | | | X | | 60 | 69 |
| 2017 | 5 | 6 | 109 | 1,6 | 68 | | | | | X | 225,6 | 24 |
| 2017 | 6 | 27 | 103 | 2,5 | 41 | | | X | | | 54,2 | 52 |
| 2017 | 7 | 23 | 120 | 1,8 | 67 | | | | X | | 86,5 | 134 |
| 2017 | 7 | 31-1 | 77 | 2 | 39 | | X | | | | 22,9 | 30 |
| 2017 | 9 | 9 | 82 | 1,1 | 75 | | | | | X | 44,7 | 66 |
| 2018 | 3 | 10-11 | 100 | NA | NA | | | | X | | 82,4 | 61 |
| 2018 | 5 | 13 | 102 | 0,8 | 128 | X | | | | | 48,1 | 20 |
| 2018 | 5 | 28 | 114 | 0,4 | 285 | | X | | | | 57,2 | NA |
| 2018 | 7 | 4 | 97 | 1,1 | 88 | | | X | | | 181,7 | 51 |
| 2018 | 7 | 13 | 64 | 2,3 | 28 | | | X | | | 118,5 | NA |
| 2018 | 7 | 16 | 150 | 4,2 | 36 | | | | X | | 104,7 | 145 |
| 2018 | 7 | 18 | 76 | 1,6 | 48 | | | | | X | 114,8 | 8 |
| 2018 | 7 | 20 | 106 | NA | NA | | | | X | | 195,1 | NA |
| 2018 | 9 | 4 | 73 | 1,1 | 66 | X | | | | | 55,6 | 10 |
| 2019 | 6 | 7 | 98 | NA | NA | | | | X | | 142,9 | 8 |
| 2019 | 6 | 13-14 | 79 | NA | NA | | | | | X | 178,6 | 45 |
| 2019 | 6 | 21 | 92 | 2,3 | 40 | | X | | | | 209,7 | 5 |
| 2019 | 7 | 7-8 | 130 | 3,9 | 33 | | X | | | | 152,1 | 99 |
| 2019 | 7 | 14 | 175 | 3,9 | 45 | | | | X | | 300,7 | 55 |
| 2019 | 10 | 7 | 92 | 0,8 | 115 | | X | | | | 30,1 | NA |
| 2019 | 10 | 23 | 103 | 1,6 | 64 | X | | | | | 30,7 | 30 |
| 2019 | 11 | 4 | 96 | 0,6 | 160 | | | | X | | 33,1 | 8 |
| 2020 | 4 | 17-18 | 66 | 1,4 | 47 | | | X | | | 86,4 | NA |
| 2020 | 6 | 7 | 96 | 2,3 | 42 | | X | | | | 57,1 | 76 |
| 2020 | 6 | 9 | 76 | 0,3 | 253 | | | X | | | 24,3 | 14 |
| 2021 | 1 | 22 | 60 | 2,9 | 21 | X | | | | | NA | 44 |
| 2021 | 4 | 30 | 82 | 0,9 | 91 | | | | X | | NA | 170 |
| 2021 | 5 | 10 | 65 | 1,7 | 38 | X | | | | | NA | NA |
| 2021 | 5 | 23 | 109 | 3,2 | 34 | | X | | | | NA | 239 |
| 2021 | 6 | 18-20 | 135 | 2,9 | 47 | X | | | | | NA | 155 |
| 2021 | 7 | 12 | 91 | 1,6 | 57 | | X | | | | NA | 16 |
| 2021 | 7 | 24 | 60 | 2,3 | 26 | | X | | | | NA | NA |
| 2021 | 8 | 9-11 | 78 | 3 | 26 | | | X | | | NA | 36 |
| 2021 | 9 | 1-4 | 75 | 2,2 | 34 | | X | | | | NA | 18 |
| 2021 | 9 | 16 | 120 | 1,7 | 71 | | | | | X | 35,7 | 267 |
| 2021 | 9 | 23-25 | 118 | 1,5 | 79 | | X | | | | 111,6 | NA |
| 2022 | 6 | 2-4 | 90 | 2,3 | 39 | X | | | | | 243 | 188 |
| 2022 | 6 | 23-26 | 95 | 1,6 | 59 | | | | | X | 142,3 | 135 |
| 2022 | 7 | 6 | 60 | 2 | 30 | | | | X | | 52,4 | NA |
| 2022 | 8 | 14 | 60 | 1,4 | 43 | | | | X | | 29,7 | 90 |
| 2022 | 8 | 17-18 | 100 | 3,6 | 28 | | X | | | | 231,9 | 117 |
| 2022 | 9 | 5 | 90 | 2,6 | 35 | | X | | | | 154,3 | 83 |
| 2022 | 9 | 14 | 80 | NA | NA | | X | | | | 164,1 | 122 |
| 2023 | 4 | 1 | 79 | 2,9 | 27 | | | X | | | 51,6 | NA |
| 2023 | 7 | 20-25 | 116 | 1,8 | 64 | X | | | | | 224,9 | 166 |
| 2023 | 8 | 4 | 79 | 1,1 | 72 | | | X | | | 126 | 38 |
| 2023 | 8 | 27 | 125 | 5,9 | 21 | | | X | | | 46,9 | 239 |
| 2023 | 9 | 18 | 82 | 1 | 82 | | | X | | | 58,2 | NA |
| 2024 | 3 | 3 | 85 | 0,9 | 94 | | | | X | | 35,6 | 62 |
| 2024 | 4 | 8-9 | 140 | 1,7 | 82 | | | | X | | 157,5 | 110 |
| 2024 | 6 | 7 | 70 | 2,2 | 32 | | | X | | | 38 | 106 |
| 2024 | 6 | 19 | 163 | 3,5 | 47 | X | | | | | 218,7 | 172 |
| 2024 | 6 | 28 | 93 | 1,3 | 72 | | | | X | | 190,2 | 108 |
| 2024 | 7 | 1 | 90 | 2 | 45 | | X | | | | 212,7 | 144 |
| 2024 | 7 | 6 | 95 | 1,4 | 68 | | X | | | | 153,3 | 102 |
| 2024 | 9 | 18 | 82 | 0,3 | 273 | | | | X | | 15,3 | 12 |
| 2025 | 3 | 21 | 82 | 1,2 | 68 | | X | | | | 40,1 | 55 |
| 2025 | 5 | 3 | 153 | 2,7 | 57 | X | | | | | 57 | 102 |
| 2025 | 5 | 19 | 83 | 2,7 | 31 | | | | X | | 107,6 | 53 |